\documentclass[10pt,journal,compsoc]{IEEEtran}
\usepackage{listings}
\usepackage[utf8]{inputenc}
\usepackage{pmboxdraw}
\usepackage{multicol}
\usepackage{pifont}
\usepackage{dcolumn}
\usepackage{bm}
\usepackage[T1]{fontenc}
\usepackage{fancyhdr}
\usepackage{verbatim}
\usepackage{bookmark}
\usepackage{changepage}
\usepackage{float}
\usepackage{mathtools}
\mathtoolsset{showonlyrefs}
\usepackage[normalem]{ulem}
\usepackage{listings}
\usepackage{fmtcount}
\usepackage{siunitx}
\usepackage{multirow}
\usepackage{relsize}
\usepackage[hyphens]{xurl}
\usepackage{hyperref}
\hypersetup{
  breaklinks,
  hidelinks,
  colorlinks=false,
}
\usepackage[hyperfirst=false,shortcuts]{glossaries}
\usepackage{xcolor}
\usepackage[nocompress]{cite}
\usepackage[pdftex]{graphicx}
  \graphicspath{{./figures/}}
  \DeclareGraphicsExtensions{.pdf,.tex}
\usepackage{tikz}
\usetikzlibrary{calc,fit,positioning,fadings,backgrounds,shapes,decorations.pathreplacing,intersections,arrows,arrows.meta}
\usepackage{gincltex}
\usepackage{amsmath}
\usepackage{array}
\usepackage[caption=false,font=footnotesize,labelfont=sf,textfont=sf]{subfig}
\usepackage[colorinlistoftodos,color=yellow!20,]{todonotes}
\usepackage{xargs}
\usepackage{soul}
\usepackage{lipsum}
\usepackage{booktabs}
\usepackage{colortbl}
\usepackage{xfp}
\usepackage{amssymb}
\usepackage{varwidth}

\colorlet{b}{black!100}
\colorlet{g}{blue}
\pgfdeclarelayer{back}
\pgfdeclarelayer{front}
\pgfsetlayers{back,main,front}

\makeatletter
\pgfkeys{%
  /tikz/on layer/.code={
    \pgfonlayer{#1}\begingroup
    \aftergroup\endpgfonlayer
    \aftergroup\endgroup
  },
  /tikz/node on layer/.code={
    \gdef\node@@on@layer{%
    \setbox\tikz@tempbox=\hbox\bgroup\pgfonlayer{#1}\unhbox\tikz@tempbox\endpgfonlayer\egroup}
    \aftergroup\node@on@layer
  },
  /tikz/end node on layer/.code={
    \endpgfonlayer\endgroup\endgroup
  }
}

\def\node@on@layer{\aftergroup\node@@on@layer}

\makeatother

\makeatletter
\def\mathcolor#1#{\@mathcolor{#1}}
\def\@mathcolor#1#2#3{%
  \protect\leavevmode
  \begingroup
    \color#1{#2}#3%
  \endgroup
}
\makeatother

\lstdefinestyle{CStyle}{
    language=C,
    keywordstyle=\ttfamily\bfseries,
    numberstyle=\footnotesize,
    basicstyle=\ttfamily\fontsize{8}{8}\selectfont,
    numbers=left,
    numberstyle=\tiny\color{darkgray},
    showstringspaces=false,
    stepnumber=1,
    numbersep=5pt,
    tabsize=2,
    captionpos=b,
    breaklines=true,
    commentstyle=\tt\color{blue},
    xleftmargin=1.5em,
    framexleftmargin=1.5em,
    linewidth=\textwidth,
    fontadjust,
    emph={barrier_m,barrier_c,line_m,line_c},
    emphstyle={\color{purple}},
}

\lstdefinestyle{JavaStyle}{
    language=Java,
    keywordstyle=\ttfamily\bfseries,
    numberstyle=\footnotesize,
    basicstyle=\ttfamily\fontsize{8}{8}\selectfont,
    numbers=left,
    numberstyle=\tiny\color{darkgray},
    showstringspaces=false,
    stepnumber=1,
    numbersep=5pt,
    tabsize=2,
    captionpos=b,
    breaklines=true,
    commentstyle=\tt\color{blue},
    xleftmargin=1.5em,
    framexleftmargin=1.5em,
    linewidth=\textwidth,
    fontadjust,
    emph={openChannel,closeChannel,log_format,handler_type,logTS,},
    emphstyle={\color{purple}},
}

\lstdefinestyle{PseudoStyle}{
    language=c++,
    keywordstyle=\ttfamily\bfseries,
    numberstyle=\footnotesize,
    basicstyle=\ttfamily\fontsize{8}{8}\selectfont,
    numbers=left,
    numberstyle=\tiny\color{darkgray},
    showstringspaces=false,
    stepnumber=1,
    numbersep=5pt,
    tabsize=2,
    captionpos=b,
    breaklines=true,
    commentstyle=\tt\color{blue},
    xleftmargin=1.5em,
    framexleftmargin=1.5em,
    linewidth=\textwidth,
    fontadjust,
    emph={throughput,duration},
    emphstyle={\color{purple}},
}

\lstdefinestyle{PythonStyle}{
    language=python,
    keywordstyle=\ttfamily\bfseries,
    numberstyle=\footnotesize,
    basicstyle=\ttfamily\fontsize{8}{8}\selectfont,
    numbers=left,
    numberstyle=\tiny\color{darkgray},
    showstringspaces=false,
    stepnumber=1,
    numbersep=5pt,
    tabsize=2,
    captionpos=b,
    breaklines=true,
    commentstyle=\tt\color{blue},
    xleftmargin=1.5em,
    framexleftmargin=1.5em,
    linewidth=\textwidth,
    fontadjust,
    emph={cjA, cjA_best, cjA_err, cmA, cmA_best, cmA_err, ratio, cyA, d_A,
    d_A_err},
    emphstyle={\color{g}},
}

\newcounter{tbdcounter}
\newcounter{todocounter}
\newcommandx{\todocount}[2][1=]{\stepcounter{todocounter}\todo[#1]{{\bf TD~\thetodocounter}: #2}}

\DeclareSIUnit \terawatthour{{\tera\watt\hour}}
\DeclareSIUnit{\million}{\text{million}}
\DeclareSIUnit{\billion}{\text{billion}}
\DeclareSIUnit{\USD}{\text{USD}}
\DeclareSIUnit{\tuple}{\text{T}}
\DeclareSIUnit{\event}{\text{E}}
\DeclareSIUnit{\request}{\text{REQ}}
\DeclareSIUnit{\ppm}{ppm}

\newcommand*{\rom}[1]{\uppercase\expandafter{\romannumeral #1\relax}}
\newcommandx{\rev}[2][1=]{\todo[linecolor=blue,backgroundcolor=blue!25,bordercolor=blue,#1]{#2}}

\newcommand\percentage[2][round-precision = 1]{
  \SI[round-mode = places,
  scientific-notation = fixed, fixed-exponent = 0,
  output-decimal-marker={.}, #1]{#2e2}{\percent}%
}

\newcommand\fnum[2][round-precision = 2]{%
  \num[round-mode = places,
  scientific-notation = fixed, fixed-exponent = 0,
  output-decimal-marker={.}, #1]{#2}%
}

\newcommand\CPINTERMAX{\SI[round-mode=places,round-precision=1]{51.20175827847414}{\micro\second}}
\newcommand\CPINTERMIN{\SI[round-mode=places,round-precision=2]{32.89531140614902}{\micro\second}}
\newcommand\CPINTRAMAX{\SI[round-mode=places,round-precision=1]{4.08428412878}{\nano\second}}
\newcommand\CPINTRAMIN{\SI[round-mode=places,round-precision=2]{5.90909}{\femto\second}}

\makeatletter
\def\mycmd{\@ifnextchar[{\@with}{\@without}}
\def\@with[#1]#2{Hello #1, have you met #2?}
\def\@without#1{Goodbye #1.}
\makeatother

\setacronymstyle{long-short}

\newglossaryentry{C++}{name={\makebox{C\hspace{-.04em}\raisebox{.1em}{\small\bf
+}\hspace{-.04em}\raisebox{.1em}{\small\bf +}}},description={C++}}
\newglossaryentry{YhSB}{name={Yahoo streaming benchmark},description={Yahoo
Streaming Benchmark}}
\newglossaryentry{TSCsl}{name={TSC client},plural={TSC clients},description={TSC client}}
\newglossaryentry{TSCma}{name={TSC server},plural={TSC servers},description={TSC server}}
\newacronym{CP}{CP}{Cloudprofiler}
\newacronym{GCE}{GCE}{Google Compute Engine}
\newacronym{ESMA}{ESMA}{European Securities and Market Authority}
\newacronym{MiFID2}{MiFID \rom{2}}{Markets in Financial Instruments Directive
\rom{2}}
\newacronym{RTT}{RTT}{round-trip time}
\newacronym{PinRTT}{PinRTT}{GPIO-Pin round-trip time}
\newacronym{MinRTT}{MinRTT}{minimum round-trip time}
\newacronym[plural=TSCs,firstplural=timestamp counters]{TSC}{TSC}{timestamp
counter}
\newacronym{NTP}{NTP}{Network Time Protocol}
\newacronym{PTP}{PTP}{Precision Time Protocol}
\newacronym{ReT}{ReT}{return-trip method}
\newacronym[plural=SPEs,firstplural=stream processing engines]{SPE}{SPE}{stream
processing engine}
\newacronym{JNI}{JNI}{Java Native Interface}
\newglossaryentry{cnfsvr}{name={configuration server},description={configuration server}}
\newacronym{PPS}{PPS}{pulse-per-second}
\newacronym{IDh}{ID handler}{identity handler}
\newacronym[plural=buffered ID handlers,firstplural=buffered identity handlers]{bIDh}{buffered ID handler}{buffered identity handler}
\newacronym[plural=PC handlers,firstplural=periodic counter handlers]{PCh}{PC handler}{periodic counter handler}
\newacronym{SPCh}{SPC handler}{single-threaded periodic counter handler}
\newacronym{MPCh}{MPC handler}{multi-threaded periodic counter handler}
\newacronym[plural=FirstLast handlers,firstplural=FirstLast
handlers]{FLh}{FirstLast handler}{FirstLast handler}
\newacronym{ReTh}{ReT handler}{return-trip handler}
\newacronym{GPIOh}{GPIO handler}{GPIO handler}
\newacronym{DG}{DG}{data generator}
\newacronym{JOF}{JOF}{Java object factory}
\newacronym{SUT}{SUT}{system-under-test}
\newacronym{APM}{APM}{application performance management}
\newacronym{JVM}{JVM}{Java virtual machine}

\glsdisablehyper

\begin{document}

\title{%
Cloudprofiler: TSC-based inter-node profiling and high-throughput
data ingestion for cloud streaming workloads
}

\author{Shinhyung~Yang, 
        Jiun~Jeong, 
        Bernhard~Scholz 
        and %
        Bernd~Burgstaller 
        \IEEEcompsocitemizethanks{\IEEEcompsocthanksitem%
          S.~Yang, %
          J.~Jeong, %
          and %
          B.~Burgstaller
          are with the Department of Computer Science, Yonsei University,
          Republic of Korea.\protect\\
        E-mail: \{shinhyung.yang,jiun.jeong,bburg\}@yonsei.ac.kr}%
        \IEEEcompsocitemizethanks{\IEEEcompsocthanksitem%
          B.~Scholz %
          is with the School of Computer Science, University of Sydney,
          Australia.\protect\\
        E-mail: bernhard.scholz@sydney.edu.au}%
}

\IEEEtitleabstractindextext{%
\begin{abstract}
\glsresetall
Real-time analytics workloads require big data stream processing engines to
process unbounded data streams at millions of events per second. However,
current streaming engines exhibit low throughput and high tuple processing
latency.
Performance engineering of streaming engines is a complex task because
they constitute distributed systems with computations orchestrated across multiple cloud nodes.
Hence a profiling technique capable of measuring time durations
at high accuracy across the nodes of a streaming engine is required.
The network time protocol 
establishes a notion of global time by syncing each client node's local clock
to one or more designated time servers. The network time protocol 
is limited to millisecond accuracy and thus unsuitable for fine-grained
performance measurements. Special-purpose
hardware devices such as satellite-derived reference clocks are not available to the guest virtual machines
of public cloud offerings and thus not viable for enhancing the accuracy of the
guest's local clock.

We propose an inter-node time duration measurement technique that obviates the
need to maintain global time. Instead, we establish a linear relation between the
timestamp counters of each pair of nodes to derive time durations where the start and end events
occur on different nodes of the
streaming framework. The precision of the relation determines the measurement
accuracy. This relation is obtained in quiescent periods of the network to
achieve accuracy in the tens of microseconds on a \SI{10}{\giga\bit\per\second}
Ethernet network. We propose a throughput-controlled data generator to reliably
determine the sustainable throughput of a streaming engine. We facilitate
high-throughput data ingestion with our concurrent object factory that moves
data tuple deserialization overhead off the critical path. The evaluation of
the proposed techniques within the Apache Storm streaming framework on the
Google~Compute~Engine shows that data ingestion increases from
\SI{700}{\kilo{}} to \SI{4.68}{\mega{}} tuples per second. Our technique
enables profiling time durations at a measurement accuracy of \CPINTERMAX{},
three orders of magnitude higher than the accuracy of NTP and one order of
magnitude higher than prior work.
\end{abstract}

}

\maketitle

\glsresetall

\IEEEdisplaynontitleabstractindextext

\IEEEpeerreviewmaketitle

\IEEEraisesectionheading{\section{Introduction}\label{sec:introduction}}

\IEEEPARstart{S}{tream} processing engines (SPEs)\glsunset{SPE}
are required to process potentially unbounded data streams at low latency and
high throughput.
The music streaming service Spotify reports that their user
event load increased from \num{1.5}~million tuples per second
(\si{\mega\tuple\per\second}) in 2016 to \SI{8.1}{\mega\tuple\per\second} in
2019~\cite{Spotify}; however, \acp{SPE} in production perform below
\SI{1}{\mega\tuple\per\second}~\cite{YahooBenchmark2016,DTTypes}.
Griffin~\textit{et al.}~\cite{Griffin2020} focus on \emph{latency-demanding}
applications that require low latency ranging from
\SIrange{1}{200}{\milli\second} and show that current data centers are
inadequate to deploy such latency-demanding applications.
A recent work~\cite{LatencyAwareSPEDeployment2020} proposes stream application
deployment strategies based on end-to-end latency that decreases communication
overhead between actors.
Hence, \acp{SPE} require performance engineering to achieve low latency and
high throughput.

\begin{figure}[t]
  \centering
  \includegraphics[width=\linewidth,trim={0 0 0 0},clip=true]
  {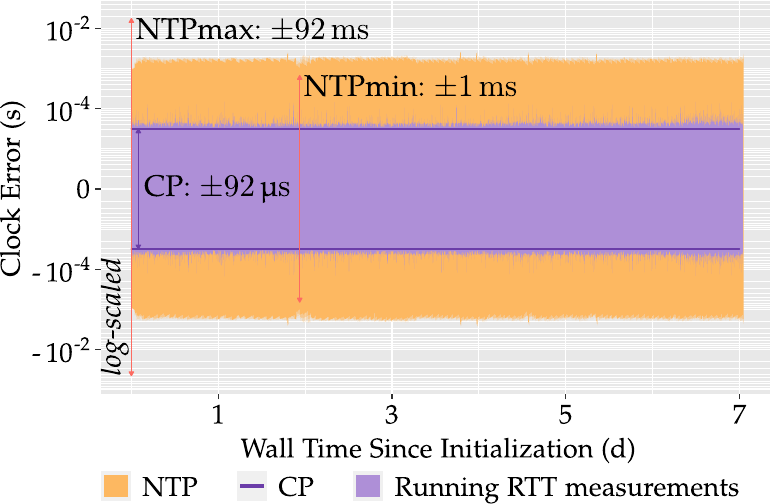}
  \caption[Error bounds of inter-node measurements: NTP vs.\ CP]
  {%
    Clock-error bounds of inter-node measurements using \ac{NTP} vs.\ \ac{CP}.
    We evaluated the clock-error bounds over the duration of a week using two
    \ac{GCE} nodes.
  }%
  \label{fig:intro-clockerr}
\end{figure}

In this paper, we propose \ac{CP}, which provides a novel measurement
methodology for latency and throughput of an \ac{SPE} (depicted in
Figures~\ref{fig:intro-clockerr} and \ref{fig:intro-throughput}) that consists
of multiple cloud nodes.

Existing inter-node duration measurement methods depend on a global time base
for time synchronization; \textit{e.g.}, a safety-critical real-time system
requires a global clock to establish a temporal order between inter-node
events~\cite{Kopetz22}.  Synchronizing to the global clock requires continuous
packets in a congested network, which decrease clock accuracy.  Performing
arithmetic operations on the unit of second further degrade the accuracy
because of the quartz-to-second conversion -- quartz has \SI{20}{\ppm} error
due to temperature and aging~\cite{Najafi2021}.  Thus, the current methods do
not meet a low-latency requirement, \textit{e.g.},\emph{``Commission Delegated
Regulation (EU) 2017/57''}~\cite{EU2017} that imposes \SI{100}{\micro\second}
accuracy between two computers.

Performance engineering of \acp{SPE} considers two metrics -- latency and
throughput~\cite{Drizzle2017, StreamVertical}.
First, latency measurement computes the elapsed time between two inter-node
events. One approach is \ac{NTP}~\cite{NTPv4}, which synchronizes each computer
clock to a single global clock within its synchronization hierarchy; it
guarantees that each synchronized clock timestamps the start and end events
with a millisecond-level accuracy.
Another approach~\cite{ReTMethod} moves the end event to the computer where the
start event occurred.
Then, the same clock that measured the start event measures the end event.
This method leverages using a single \ac{TSC} for both start and end events
with the longest returning duration determining its accuracy bound. In
the remainder of this paper, we call this approach the \ac{ReT}.
Second, throughput measurement counts the number of processed tuples per unit
time. One approach~\cite{Karimov18} employs external data generators, lowering
the generation rate until no backpressure occurs in the \ac{SPE}. Researchers
distinguish vertical scalability (scale-up)~\cite{StreamVertical} from
horizontal scalability (scale-out)~\cite{Theodolite} and employ parameter
auto-tuning to increase the throughput~\cite{TuningSurvey, Tuning:Bilal,
Tuning:Storm}.
Performance engineering an \ac{SPE} is a challenging
task because an \ac{SPE} loses performance by partitioning a streaming
application into its computational actors and scheduling them
across cloud nodes; consequently, the actors require inter-process and
inter-node communications, leading to latency and throughput degradation.

Performance evaluation of an \ac{SPE} is a complex task~\cite{APM2017} because
its architecture is multi-tiered (\ac{JVM} and machine code), multi-layered
(hardware, host OS, hypervisor, and guest OS), and consists of multiple cloud
nodes.  To evaluate an \ac{SPE}, \ac{CP} employs several strategies.  First,
\ac{CP} executes as machine code, separating itself from the \ac{JVM} that runs
the \ac{SPE}; \ac{CP} alleviates the \ac{JVM} overhead from running its logging
facility that collects, processes, and stores data. Second, \ac{CP} employs
dedicated log compression threads to saturate the maximum disk write bandwidth.
Third, \ac{CP} eliminates the overhead of recompiling and uploading the
\ac{SPE} to multiple cloud nodes using the \ac{cnfsvr} that selects a specific
logging type for a particular instrumentation location in the code at runtime.

Fig.~\ref{fig:intro-clockerr} highlights \ac{NTP}'s shortcomings for measuring
an inter-node duration. The result confirms that an \ac{NTP}-synchronized clock
is \SI{1}{\milli\second} accurate at best, which is inadequate for inter-node
duration measurement.
Related works point out that \Ac{NTP} provides an accuracy of
\SI{35}{\milli\second} on a personal computer~\cite{NTP35ms} and
\SI{1}{\milli\second} in a controlled environment~\cite{NTPInWindowsServer2016,
FineTuneNTPInWindowsServer2016}.
Dongen~\textit{et al.}~\cite{Eval2020} confirm the problem with \ac{NTP}'s
accuracy;  they could not measure workload latency shorter than
\SI{10}{\milli\second} across nodes with \ac{NTP} synchronization.
Prior work~\cite{Tuning:Bilal} uses \ac{NTP}-synchronized clocks and reports
inter-node durations smaller than \SI{50}{\milli\second}, but the results lack
the clock error information, which is crucial for validating their measurement.
Weber~\textit{et al.}~\cite{weber2017availability} point out that Ethereum, a
mainstream Blockchain lacks a clock synchronization accuracy requirement for
participating nodes, which is essential for measuring system's availability.
Their proposed timing measurement is only accurate to \SI{1}{\second}.

Fig.~\ref{fig:intro-clockerr} validates \ac{CP} as a more accurate inter-node
duration measurement method than \ac{NTP}. \ac{CP} uses \ac{MinRTT} to relate
two inter-node \ac{TSC} values with a microsecond-level error bound. A \ac{TSC}
is nonstop, which resides in a CPU and runs at a constant
rate; therefore, it is a qualified source for measuring single-node
durations~\cite{PAPI2000}. However, it cannot measure inter-node durations
because \acp{TSC} on different computers boot at different times and run at
different rates. \Ac{CP} presents a \emph{reference \ac{TSC}}, pursuing the
advantage of using a single \ac{TSC} for duration measurements. Two \ac{MinRTT}
measurements establish a linear relation between the \acp{TSC} of two nodes,
and one \ac{TSC} becomes the reference \ac{TSC} that provides the standard unit
time for inter-node durations. Fig.~\ref{fig:intro-clockerr} shows that the
\ac{MinRTT} error bound is \SI{92}{\micro\second},
and \ac{NTP}'s accuracy is bounded by the interval
$[1, 92]$~\si{\milli\second}.
\Ac{MinRTT} is more accurate than \ac{NTP}, up to three orders of magnitude.

\begin{figure}[t]
  \centering
  \includegraphics[width=\linewidth,trim={0 0 0 0},clip=true]
  {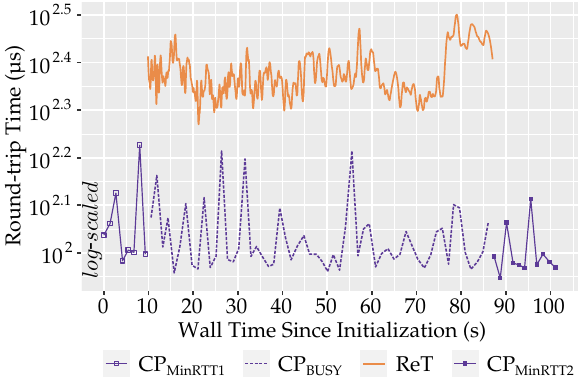}
  \caption[Accuracy of the CP and ReT measurement schemes]
  {%
    Benchmark execution starts \SI{10}{\second} after system initialization.
    \Ac{CP} is accurate to the \ac{MinRTT} measured before the benchmark
    execution, whereas \ac{ReT} is accurate to the maximum \ac{RTT} measured in
    the congested network.
  }%
  \label{fig:intro-CPvsReT}
\end{figure}

Fig.~\ref{fig:intro-CPvsReT} introduces \Ac{ReT}, another approach to measuring
an inter-node duration between start and end events~\cite{ReTMethod}.  \Ac{ReT}
resends all end events to the starting node and timestamps them as they arrive.
It utilizes a single \ac{TSC} for both events, effectively eliminating the
\ac{NTP} synchronization requirement.  However, \ac{ReT} always incurs an extra
duration for resending end events back to the starting node, and the maximum
extra duration determines \ac{ReT}'s error bound.  We measure \ac{ReT}'s error
bound by retrieving the \ac{RTT} of the returned end events.
Fig.~\ref{fig:intro-CPvsReT} compares the \ac{RTT} results of \ac{CP} and
\ac{ReT}. \Ac{ReT} yields higher \acp{RTT} (data series ``$\text{ReT}$'')
because it competes for the network resource with the ongoing experiment.
\Ac{CP} exploits quiescent periods -- \ac{MinRTT} runs before and after the
experiment (data series ``$\text{CP}_\text{MinRTT1}$'' and
``$\text{CP}_\text{MinRTT2}$''). As a reference point, we also ran \ac{MinRTT}
alongside the ongoing experiment (data series ``$\text{CP}_\text{BUSY} $''),
which also performed better than \ac{ReT}.

\Ac{CP} has the following advantages.
  First, it 
  synchronizes the \acp{TSC} in a quiescent period and thus
  uses the best network condition rather than the current condition,
  unlike~\cite{ReTMethod} or \cite{OneWay}.
  Second, it has no dependency on the global clock, unlike \ac{NTP}.
  Third, it provides a strict, analytical error bound rather than statistical
  accuracy~\cite{TenMicroseconds}.
  Lastly, it provides off-the-shelf availability on the cloud; it does not
  require peripheral hardware support such as GPS receivers or
  \ac{PTP}~\cite{PTPv2} devices unavailable on commodity cloud
  nodes.

\begin{figure}[t]
  \centering
  \includegraphics[width=\linewidth,trim={0 0 0 0},clip=true]
  {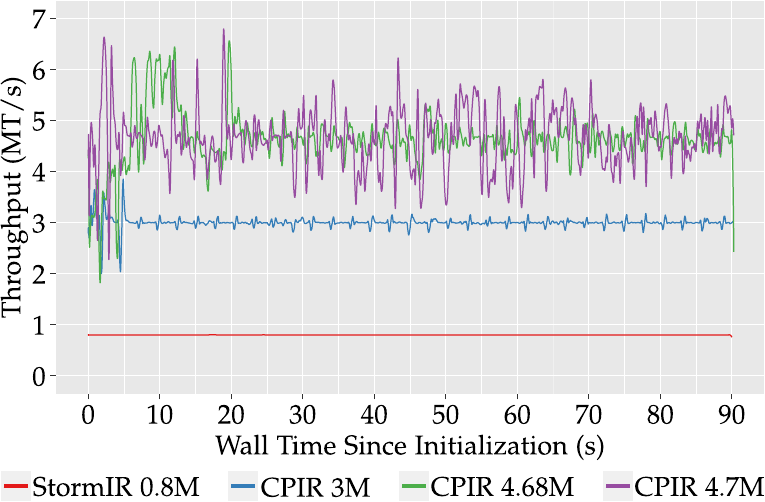}
  \caption[Ingestion rate of Apache Storm, unoptimized and optimized]
  {%
    Ingestion rates of an unoptimized Storm (StormIR) and Storm with our
    optimization (CPIR) at a constant throughput of
    \SI{3}{\mega\tuple\per\second}, \SI{4.68}{\mega\tuple\per\second}, and
    \SI{4.7}{\mega\tuple\per\second}.
  }%
  \label{fig:intro-throughput}
\end{figure}

\begin{figure}[t]
  \centering
  \includegraphics[width=\linewidth,trim={0 0 0 0},clip=true]
  {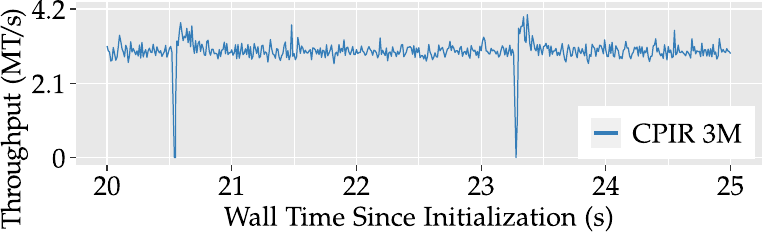}
  \caption[Java garbage collection cycles with Apache Storm]
  {%
    Our throughput optimizations for Storm require the Java garbage collector
    to reclaim memory more frequently.
  }%
  \label{fig:intro-throughput-magnified}
\end{figure}

Figures~\ref{fig:intro-throughput} and~\ref{fig:intro-throughput-magnified}
show our \ac{SPE} throughput optimization result using our \emph{maximum
throughput measurement method}. An \ac{SPE} receives unbounded data streams and
computes them at the same time.  \emph{Sustainable throughput}~\cite{Karimov18}
is a metric for \ac{SPE} throughput that continuously lower the emission rate
until no tuple backpressure occurs. We propose a \emph{maximum throughput
measurement method } that experimentally evaluates the maximum throughput of an
\ac{SPE} by increasing the emission rate with a \ac{C++} throughput-controlled
emission loop and a \ac{JOF}. If the \ac{SPE} does not drop any tuples at a
constant emission rate, we repeat the experiment with an increased emission
rate. We do this iteratively until the \ac{SPE} exhibits a tuple drop, and we
pick the last successful throughput as the maximum throughput.

In Fig.~\ref{fig:intro-throughput}, we show the result of the maximum
throughput measurement method on an unoptimized Storm \ac{SPE} (StormIR) and
the ingestion rate of a Storm \ac{SPE} with our optimization (CPIR).  Using our
method, the maximum sustainable throughput of CPIR is
\SI{4.7}{\mega\tuple\per\second}.  Fig.~\ref{fig:intro-throughput-magnified}
magnifies the fluctuation at the rate of \SI{3}{\mega\tuple\per\second}; we
discovered that it is caused by the Java garbage collector when there is no
space for new allocations, especially deserialized Java strings. This paper
makes the following contributions.
\begin{enumerate}
  \item We propose \ac{CP} inter-node measurement method that is accurate to
    \SI{40}{\micro\second} across two nodes in the cloud.

  \item We measured the maximum error~bounds of Ethernet-based measurement
    methods on 30~cloud nodes. Our experiment proves that the \ac{CP} method is
    the most accurate on commodity cloud nodes.

  \item We measured an \ac{SPE} can ingest tuples at constant throughput of
    \SI{4.7}{\mega\tuple\per\second} at the maximum.

  \item We propose a \ac{CP} cloud profiling framework, written in \ac{C++}11
    for native platforms, providing platform-independent APIs that log a tuple
    with or without compression in \SI{67.5}{\nano\second} and
    \SI{80.7}{\nano\second}, respectively.
\end{enumerate}

The remainder of the paper is structured as follows.
In Section~\ref{s:bg}, we discuss existing measurement methods, including the
\ac{NTP}-based measurement method and the \ac{ReT}-based measurement
method~\cite{ReTMethod}.
Section~\ref{s:cp}, describes the proposed method and its accuracy using
the \ac{MinRTT} of two \acp{TSC} on each computer.
In Section~\ref{s:dg}, we discuss our maximum throughput measurement method
that correctly evaluates the throughput performance of an \ac{SPE} using
throughput-controlled emission loops and the \acl{JOF}.
In Section~\ref{s:val}, our proposed validations are explained, including
\ac{ReT} validation and \ac{TSC} validation.
Section~\ref{s:exp} describes our experimental setups, implementation details,
and conducting details of the streaming benchmark. Then we evaluate the
experimental results.
Section~\ref{s:rel} introduces prior works that precede our inter-node
measurement methodology and compares the contributions.
We draw our conclusions in Section~\ref{sec:concl}.

\section{Background\label{s:bg}}

\subsection{NTP-based Measurement and Clock Error}
Traditionally, and until now, most time duration measurement in Java
applications -- in a single node or across nodes -- has been done by comparing
the \emph{System.currentTimeMillis()} timestamps. There has been no awareness
of \emph{\acf{NTP}} synchronization, which uses a hierarchy of computers
synchronized to a reference clock~\cite{mills2017ntp}.

{\setlength\tabcolsep{3pt}
\renewcommand{\arraystretch}{1.5}
\begin{table}[tbh]
  \caption[NTP System Variables]
  {%
    \ac{NTP} System Variables
  }%
  \begin{center}
  \begin{tabular}{ c p{0.83\linewidth} }
  \toprule
  Variable & Name and Description \\
  \midrule
  $\Delta$  & \emph{root\_delay} is the total sum of round-trip delay that
  should compensate network asymmetries inherent to all packet delays from a
  local computer to the primary server (stratum-1) which has a locally attached
  reference clock where all downstream servers and clients are ultimately
  synchronized to.
              \\
  $E$       & \emph{root\_dispersion} is the accumulated dispersion over the
  network from the primary server, where dispersion $\epsilon$ represents the
  maximum error inherent in the measurement at each level between two
  consecutive servers.
              \\
  $\Theta$  & \emph{system\_time\_offset} is the remaining offset to be slewed
  off from the local clock.
              \\
  \bottomrule
  \end{tabular}
  \end{center}
  \label{tab:ntp}
\end{table}
}%

\begin{equation}\label{eq:ntp1}
  \left|\theta\right| \leq \Theta + E + \left(0.5 * \Delta\right)
\end{equation}

An error bound is the minimum or the maximum error of a measured quantity.
Inequality~\eqref{eq:ntp1} states the maximum error bound for the
offset~$\theta$ of the time-server clock relative to a computer's local clock,
as specified in the \ac{NTP}v4 standard document~\cite{NTPv4}. An \ac{NTP}
client polls a sample for each unit time and stores the sample in a list, then
selects the best sample that chooses \emph{root\_dispersion} $E$ and
\emph{root\_delay} $\Delta$ for wall-clock synchronization. It adjusts the
clock at a maximum of \SI{0.5}{\milli\second\per\second} (\textit{i.e.},
adjusting one second requires \SI{2000}{\second})~\cite{NTPWeb} toward the
global clock, and \emph{system\_time\_offset} $\Theta$ denotes the current
offset of the local clock.

\subsection{Return-trip Method~(ReT) Event Occurrence\label{ssec:ret_method}}

\begin{figure}[tbh]
  \centering
  \includegraphics[width=\linewidth,trim={0 0 0 0},clip=true]
  {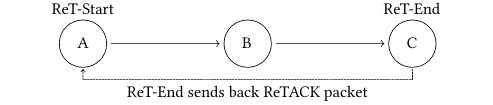}
  \caption[Return-trip measurement scheme]
  {%
    \Ac{ReT} method: an arrival of the end event in Node~$C$ sends back an
    ReTACK to Node~$A$ where the start event occurred. On its arrival in
    Node~$A$, the method measures the duration between the start and the end
    event. The method uses the \ac{TSC} of $A$ as the single \ac{TSC} for both
    the start and the end event.
  }%
  \label{fig:method_ret}
\end{figure}

Off-the-shelf cloud nodes employ \ac{NTP} to synchronize their clocks to a
global reference clock.  Assume a tuple processing ${A}\to{B}\to{C}$ as
illustrated in Fig.~\ref{fig:method_ret}. A duration starts in Node~$A$ at time
${T}_{1}$ and ends in Node~$C$ at time ${T}_{2}$.  Note that ${T}_{1}$ and
${T}_{2}$ occurs on nodes~$A$ and $C$, respectively; using time duration
${T}_{2}-{T}_{1}$ introduces a millisecond or higher error~bound.

\Ac{ReT}~\cite{ReTMethod} eliminates the problem of \ac{NTP} that incurs
millisecond-level error bound by reading the global clock within a hierarchy of
\ac{NTP} servers, denoted as a root delay $\Delta$ in Inequality~\ref{eq:ntp1}.
\Ac{ReT} avoids the use of \ac{TSC} values generated on different computers. To
measure the latency of ${A}\to{B}\to{C}$, $C$ sends back an \emph{ReTACK} to
$A$, which arrives at ${T}_{3}$. Note that ${T}_{1}$ and ${T}_{3}$ are
generated on $A$; \ac{ReT} replaces the the time duration ${T}_{2}-{T}_{1}$
with the time duration ${T}_{3}-{T}_{1}$. \Ac{ReT} eliminated the inaccuracy of
\ac{NTP} at the cost of the inaccuracy imposed by the ReTACK packet transfer
time.  We discuss the detailed implementation and its validation in
Section~\ref{exp:ret}.

\section{TSC-based Inter-node Measurement\label{s:cp}}

\newcommand{\Event}[1]{$e_{#1}$}
\newcommand{\Tick}[4][black]{$\tick[#1]{#2}{#3}{#4}$}
\newcommand{\tick}[4][black]{\mathcolor{#1}{{#2}_{#3}^{#4}}}
\newcommand{\Tk}[3][black]{$\tk[#1]{#2}{#3}$}
\newcommand{\tk}[3][black]{\tick[#1]{c}{#2}{#3}}
\newcommand{\Er}[3][black]{$\er[#1]{#2}{#3}$}
\newcommand{\er}[3][black]{\error[#1]{c}{#2}{#3}}
\newcommand{\Best}[3][black]{$\best[#1]{#2}{#3}$}
\newcommand{\best}[3][black]{\mathcolor{#1}{{\hat{c}}_{#2}^{#3}}}
\newcommand{\DurBest}[1]{$\durBest{#1}$}
\newcommand{\durBest}[1]{{\hat{\mathit{d}}}_{\mathit{#1}}}
\newcommand{\Error}[4][g]{$\error[#1]{#2}{#3}{#4}$}
\newcommand{\error}[4][g]{\delta{\mathcolor{#1}{{#2}_{#3}^{#4}}}}
\newcommand{\DurErr}[1]{$\durErr{#1}$}
\newcommand{\durErr}[1]{{\delta{\mathit{d}}}_{\mathit{#1}}}
\newcommand{\TSC}[1]{${C}^{#1}$}
\newcommand{\period}[1]{$\mathit{period}\left(C^{#1}\right)$}
\newcommand{\Ratio}[2]{$\ratio{#1}{#2}$}
\newcommand{\ratio}[2]{\mathit{ratio}\left(C^{#1},C^{#2}\right)}
\newcommand{\RatioNew}[2]{$\ratioNew{#1}{#2}$}
\newcommand{\ratioNew}[2]{\mathit{ratio}({#1},{#2})}
\newcommand{\RatioSym}[3][g]{$\ratioSym[#1]{#2}{#3}$}
\newcommand{\ratioSym}[3][g]{\mathcolor{#1}{{\mathit{r}}_{#3}^{#2}}}
\newcommand{\RatioFrac}[4]{$\ratioFrac{#1}{#2}{#3}{#4}$}
\newcommand{\ratioFrac}[4]{\left|\frac{\tk[g]{#1}{#3}-\tk[g]{#2}{#3}}{\tk{#1}{#4}-\tk{#2}{#4}}\right|}
\newcommand{\RFracII}[5][g]{$\rFracII[#1]{#2}{#3}{#4}{#5}$}
\newcommand{\rFracII}[5][g]{\frac{\left|\tk[#1]{#2}{#4}-\tk[#1]{#3}{#4}\right|}{\left|\tk{#2}{#5}-\tk{#3}{#5}\right|}}
\newcommand{\RFracNoAbs}[5][g]{$\rFracNoAbs[#1]{#2}{#3}{#4}{#5}$}
\newcommand{\rFracNoAbs}[5][g]{\frac{\tk[#1]{#2}{#4}-\tk[#1]{#3}{#4}}{\tk{#2}{#5}-\tk{#3}{#5}}}
\newcommand{\RFracSym}[5][g]{$\rFracSym[#1]{#2}{#3}{#4}{#5}$}
\newcommand{\rFracSym}[5][g]{\frac{\durSym[#1]{#2}{#3}{#4}}{\durSym{#2}{#3}{#5}}}
\newcommand{\RFSymII}[4]{$\rFSymII{#1}{#2}{#3}{#4}$}
\newcommand{\rFSymII}[4]{\durSym[g]{#1}{#2}{#3}/\durSym{#1}{#2}{#4}}

\newcommand{\DurFunc}[2]{$\durFunc{#1}{#2}$}
\newcommand{\durFunc}[2]{\mathit{duration}({#1},{#2})}
\newcommand{\durSymVII}[3]{d_{{{#1},{#2}}}^{#3}}
\newcommand{\durSymII}[3]{d_{{[{#1}..{#2}]}}^{#3}}
\newcommand{\durSymIII}[3]{d_{{{#1}..{#2}}}^{#3}}
\newcommand{\DurSym}[4][black]{$\durSym[#1]{#2}{#3}{#4}$}
\newcommand{\durSym}[4][black]{\mathcolor{#1}{d_{{[{#2},{#3}]}}^{#4}}}
\newcommand{\DurSymErr}[4][g]{$\durSymErr[#1]{#2}{#3}{#4}$}
\newcommand{\durSymErr}[4][g]{\mathcolor{#1}{\delta{d_{{[{#2},{#3}]}}^{#4}}}}
\newcommand{\DurSymV}[3]{\durSymV{#1}{#2}{#3}}
\newcommand{\durSymV}[3]{d_{{\overline{{#1}{#2}}}}^{#3}}
\newcommand{\durSymVI}[3]{d_{{\left|{#1}-{#2}\right|}}^{#3}}
\newcommand{\Duration}[6]{$\duration{#1}{#2}{#3}{#4}{#5}{#6}$}
\newcommand{\duration}[6]{d_{(#1_{#2}^{#3},#4_{#5}^{#6})}}
\newcommand{\NewDuration}[1]{$\newDuration{#1}$}
\newcommand{\newDuration}[1]{\mathit{d}_\mathit{#1}}
\newcommand{\DurSingle}[3]{$\durSingle{#1}{#2}{#3}$}
\newcommand{\durSingle}[3]{d_{\left(#1,#2\right)}^{#3}}
\newcommand{\PartDerv}[2]{$\partDerv{#1}{#2}$}
\newcommand{\partDerv}[2]{\frac{\partial{#1}}{\partial{#2}}}
\newcommand{\PartDeII}[2]{$\partDeII{#1}{#2}$}
\newcommand{\partDeII}[2]{\partial{#1} / \partial{#2}}
\newcommand{\Var}[2][black]{$\var[#1]{#2}$}
\newcommand{\var}[2][black]{\mathcolor{#1}{#2}}
\newcommand{\MinRTT}[1]{$\minRTT{#1}$}
\newcommand{\minRTT}[1]{\mathit{MinRTT}_{#1}}
\newcommand{\Freq}[1]{$\freq{#1}$}
\newcommand{\freq}[1]{\mathit{f}^{#1}}
\newcommand{\Node}[1]{Node~$#1$}
\newcommand{\Cmb}[3][']{$\cmb[#1]{#2}{#3}$}
\newcommand{\cmb}[3][']{\left({#2} \cdot {#3}\right)#1}
\newcommand{\cp}[1]{Scenario~#1}
\newcommand{\CP}[1]{\cp{#1}}
\newcommand{\Wallclock}[1]{$\wallclock{#1}$}
\newcommand{\wallclock}[1]{\mathit{wallclock}\left(#1\right)}

\subsection{Reference TSC\label{sec:refTSC}}
The \Ac{CP} method enables measuring inter-node time durations at a higher
accuracy than existing approaches.  Two characteristics of a \ac{TSC} make a
unique clock for individual nodes: (1) nonstop and (2) constant; a nonstop
\ac{TSC} does not stop regardless of the CPU's current state such as C-States;
a constant \ac{TSC} counts up at a constant rate regardless of the CPU's
clock frequency changes.
A \ac{TSC} on a typical \ac{GCE} node provides an accuracy of
\SI{0.45}{\nano\second} per cycle for measuring time durations on a single
node~\cite{GCECPUs}.
PAPI is ubiquitous for single-node performance engineering~\cite{PAPI2000};
however it was not available for the multi nodes because of unrelated \acp{TSC}
across nodes.
Thus, we introduce \emph{reference \ac{TSC}} -- the single \ac{TSC} for
measuring and comparing time durations across nodes.
It requires \emph{translating} a node-specific \ac{TSC} value to a reference
\ac{TSC} value; to accomplish this, we introduce \emph{\ac{TSC} Ratio}, a
linear relationship between two \ac{TSC} durations on two nodes, measured
within the same time duration marked by two \acp{MinRTT}.
\begin{figure}[t]
  \centering
    \includegraphics[width=\linewidth,trim={0 0 0 0},clip=true]
    {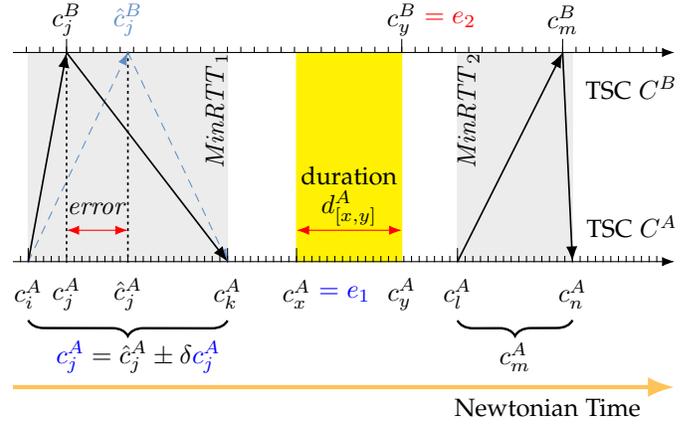}
    \caption[Inter-node duration measurement scheme]
    {%
      Measurement of an inter-node duration between
      event~\Event{1}~(\Tk{x}{A}) on Node~A and
      event~\Event{2}~(\Tk{y}{B}) on Node~B:
      The \Ac{CP} method uses the ratio of elapsed \ac{TSC} durations between
      nodes~$A$ and $B$.
    }%
    \label{fig:method_cp_ratio}
\end{figure}
Fig.~\ref{fig:method_cp_ratio} depicts two \acp{TSC} and two \acp{MinRTT}.
\MinRTT{1} and \MinRTT{2} mark the start and the end of the time
durations~\DurSym{j}{m}{A} and \DurSym{j}{m}{B}.
Using the variables in the example measurement in
Fig.~\ref{fig:method_cp_ratio}, we introduce two notations:
\RatioSym[black]{A}{B} and \DurSym{j}{m}{A}.  Ratio~\RatioSym[black]{A}{B} is a
ratio between two nodes, $A$ and $B$. Time duration~\DurSym{j}{m}{A} represents
a time duration between two \ac{TSC} values \Tk{j}{A} and \Tk{m}{A}.
Eq.~\eqref{eq:sym_cp_ratio} holds for a ratio between two \acp{TSC} on nodes~$A$ and $B$:
\begin{align}\label{eq:sym_cp_ratio}
  \ratioSym[black]{A}{B}
  = \rFracII[black]{j}{m}{A}{B}
  = \rFracSym[black]{j}{m}{A}{B}
\end{align}

\emph{Propagation of Error}.
\Ac{TSC} ticks \Tk{j}{A} and \Tk{m}{A} are approximations of \ac{TSC} ticks
\Tk{j}{B} and \Tk{m}{B}; the \acl{RTT} of each \ac{MinRTT} measurement bounds
their errors: we know that \Tk{j}{B} happened between \Tk{i}{A} and \Tk{k}{A},
but we do not know the temporal order between \Tk{j}{B} and \Best{j}{A};
\textit{i.e.}, there exists an $\mathit{error}$ between \Tk{j}{B} and
\Best{j}{A}.
A symmetric \ac{RTT} between nodes~$A$ and $B$ yields \Best{j}{B}; however,
network asymmetries always results in a \ac{TSC} tick such as \Tk{j}{B}, which
bounds to a minimum of \Tk{i}{A} and a maximum of \Tk{k}{A}; thus, \Tk{j}{B}
and \Best{j}{A} are incomparable.
Using this information, we can declare \Tk{j}{A} in the standard
error form~\cite[Eq.~(2.3)]{ErrorAnalysis}.
\begin{gather}\label{eq:StdErrForm}
  \tk[g]{j}{A} = \best{j}{A} \pm \er{j}{A}
  \end{gather}
The measured value \Tk[black]{j}{A} comprises of two parts: \Best{j}{A}
and \Er[black]{j}{A}. We take an example from
\MinRTT{1} in Fig.~\ref{fig:method_cp_ratio} that starts and ends at \Tk{i}{A} and
\Tk{k}{A}. Because we cannot determine \Tk{j}{A}, we use \Best{j}{A}, on the
halfway between \Tk{i}{A} and \Tk{k}{A}, as the 
best estimate of \Tk{j}{A}; we use \Er[black]{j}{A}, half the duration
between \Tk{i}{A} and \Tk{k}{A}, as the uncertainty that bounds the error of
\Tk{j}{A}.

To aid in understanding the propagation of errors, we will use a specific color
($\mathcolor{g}{\bullet}$) for \ac{TSC} values with uncertainty in mathematical
expressions in the current section, as shown in equations~\eqref{eq:StdErrForm}.

\subsection{Conversion of TSC values between TSC domains\label{sec:trans_val}}
A \ac{TSC} value is native to the node where it occurs; \acp{TSC} on different
nodes have different \ac{TSC} frequencies and they boot at different time.
time. Our \ac{CP} method incorporates a \emph{ratio} in the translation to make
a \ac{TSC} value available on another node.  We show the translation process
using the measurement example in Fig.~\ref{fig:method_cp_ratio}. We translate
\Tk{y}{B} to \Tk[black]{y}{A} using the ratio~\RatioSym[black]{A}{B}.
\begin{IEEEeqnarray*}{rCl}
  \tk[g]{y}{A}
  &=& \tk[g]{j}{A} + \ratioSym{A}{B}\left(\tk{y}{B} - \tk{j}{B}\right)
  \\
  &=& \tk[g]{j}{A} + \rFracII{j}{m}{A}{B}\left(\tk{y}{B} - \tk{j}{B}\right)
\end{IEEEeqnarray*}
We can eliminate the absolute value signs because $\tk{j}{A}<\tk{m}{A}$ and
$\tk{j}{B}<\tk{m}{B}$.
\begin{IEEEeqnarray}{rCl}\label{eq:cyA}
  \tk[g]{y}{A}
  &=& \tk[g]{j}{A} + \frac{\tk[g]{m}{A}-\tk[g]{j}{A}}{\tk{m}{B}-\tk{j}{B}}
  \left(\tk{y}{B} - \tk{j}{B}\right)
\end{IEEEeqnarray}
Calculating \Tk[black]{y}{A} involves \Tk[black]{m}{A} and \Tk[black]{j}{A}:
values with uncertainties~\Er{m}{A} and \Er{j}{A}; \textit{i.e.},
\Tk[black]{y}{A} also incurs an uncertainty~\Er{y}{A}. We can calculate it
using the error propagation rule~\cite[Eq.~(3.48)]{ErrorAnalysis}.
\begin{IEEEeqnarray}{rCl}\label{eq:err_prop_rule}
  \var[black]{\delta{q}} \leq
  \left|\partDerv{\var[g]{q}}{x}\right|\var[b]{\delta{x}} + \ldots +
  \left|\partDerv{\var[g]{q}}{z}\right|\var[b]{\delta{z}}
\end{IEEEeqnarray}
We calculate the uncertainty \Er{y}{A} using Eq.~\eqref{eq:err_prop_rule}.
\begin{IEEEeqnarray}{rCl}\label{eq:cyAerr}
  \er{y}{A}
  &=& \left|\partDerv{\tk[g]{y}{A}}{\tk[g]{j}{A}}\right| \er{j}{A} +
  \left|\partDerv{\tk[g]{y}{A}}{\tk[g]{m}{A}}\right| \er{m}{A}
\end{IEEEeqnarray}
To evaluate the partial derivatives of
Eq.~\eqref{eq:cyA}, we rearrange the equation in terms of \Tk[black]{m}{A} and
\Tk[black]{j}{A}.
\begin{IEEEeqnarray*}{rCl}
  \tk[g]{y}{A}
  &=& \tk[g]{j}{A} + \frac{\tk[g]{m}{A}-\tk[g]{j}{A}}{\tk{m}{B}-\tk{j}{B}}
  \left(\tk{y}{B} - \tk{j}{B}\right)
  \\
  &=& \tk[g]{j}{A} + \frac{\tk{y}{B} - \tk{j}{B}}{\tk{m}{B}-\tk{j}{B}}
  \left(\tk[g]{m}{A}-\tk[g]{j}{A}\right)
  \\
  &=& \tk[g]{j}{A}
  + \frac{\tk{y}{B} - \tk{j}{B}}{\tk{m}{B}-\tk{j}{B}}\tk[g]{m}{A}
  - \frac{\tk{y}{B} - \tk{j}{B}}{\tk{m}{B}-\tk{j}{B}}\tk[g]{j}{A}
\end{IEEEeqnarray*}
Holding \Tk[black]{m}{A} constant, we differentiate it with respect to
\Tk[black]{j}{A}.
\begin{IEEEeqnarray*}{rCl}
  \partDerv{\tk[g]{y}{A}}{\tk[g]{j}{A}}
  &=& \partDerv{}{\tk[g]{j}{A}}\left(\tk[g]{j}{A}\right)
  + \partDerv{}{\tk[g]{j}{A}}
    \left(\frac{\tk{y}{B} - \tk{j}{B}}{\tk{m}{B}-\tk{j}{B}}\tk[g]{m}{A}\right)
  \\ &&
  - \partDerv{}{\tk[g]{j}{A}}
    \left(\frac{\tk{y}{B} - \tk{j}{B}}{\tk{m}{B}-\tk{j}{B}}\tk[g]{j}{A}\right)
  \\ &=&
  1 + 0 - \frac{\tk{y}{B} - \tk{j}{B}}{\tk{m}{B}-\tk{j}{B}}
  \\ &=&
  1 - \frac{\tk{y}{B} - \tk{j}{B}}{\tk{m}{B}-\tk{j}{B}}
\end{IEEEeqnarray*}
Likewise, holding \Tk[black]{j}{A} constant, we differentiate it with respect
to \Tk[black]{m}{A}.
\begin{IEEEeqnarray*}{rCl}
  \partDerv{\tk[g]{y}{A}}{\tk[g]{m}{A}}
  &=& \partDerv{}{\tk[g]{m}{A}}\left(\tk[g]{j}{A}\right)
  + \partDerv{}{\tk[g]{m}{A}}
    \left(\frac{\tk{y}{B} - \tk{j}{B}}{\tk{m}{B}-\tk{j}{B}}\tk[g]{m}{A}\right)
  \\ &&
  - \partDerv{}{\tk[g]{m}{A}}
    \left(\frac{\tk{y}{B} - \tk{j}{B}}{\tk{m}{B}-\tk{j}{B}}\tk[g]{j}{A}\right)
  \\ &=&
  0 + \frac{\tk{y}{B} - \tk{j}{B}}{\tk{m}{B}-\tk{j}{B}} + 0
  \\ &=&
  \frac{\tk{y}{B} - \tk{j}{B}}{\tk{m}{B}-\tk{j}{B}}
\end{IEEEeqnarray*}
Now we can calculate the total error Eq.~\eqref{eq:cyAerr}.
We use the largest error \Var[b]{e} for this calculation.
\begin{IEEEeqnarray}{C}\label{eq:chooseMax}
\mathit{max}(\er{j}{A}, \er{m}{A}) = \var[b]{e}
\end{IEEEeqnarray}
We remove absolute value signs because we know $\tk{j}{B} < \tk{y}{B} <
\tk{m}{B}$.
\begin{IEEEeqnarray*}{rCl}
\er{y}{A} &=&
\left|\partDerv{\tk[g]{y}{A}}{\tk[g]{j}{A}}\right| \var[b]{e} +
\left|\partDerv{\tk[g]{y}{A}}{\tk[g]{m}{A}}\right| \var[b]{e}
\\ &=&
\left|1 - \frac{\tk{y}{B} - \tk{j}{B}}{\tk{m}{B} - \tk{j}{B}}\right| \var[b]{e}
+ \left|\frac{\tk{y}{B} - \tk{j}{B}}{\tk{m}{B} - \tk{j}{B}}\right| \var[b]{e}
\\ &=&
\left(1 - \frac{\tk{y}{B} - \tk{j}{B}}{\tk{m}{B} - \tk{j}{B}}\right) \var[b]{e}
+ \left(\frac{\tk{y}{B} - \tk{j}{B}}{\tk{m}{B} - \tk{j}{B}}\right) \var[b]{e}
\\ &=&
\left(1 + \frac{\left(\tk{y}{B} - \tk{j}{B}\right) - \left(\tk{y}{B} - \tk{j}{B}\right)}{\tk{m}{B} - \tk{j}{B}}\right) \var[b]{e}
\\ &=&
\var[b]{e}
\end{IEEEeqnarray*}
Using the standard error form, we can rewrite the final result:
\begin{IEEEeqnarray*}{rCl}
  \tk[g]{y}{A} &=&
  \best{y}{A} \pm \er{y}{A}
  \\ &=&
  \best{j}{A} + \frac{\best{m}{A}-\best{j}{A}}{\tk{m}{B}-\tk{j}{B}}
  \left(\tk{y}{B} - \tk{j}{B}\right) \pm \var[b]{e}
\end{IEEEeqnarray*}

\subsection{Translation of a TSC duration\label{sec:trans_dur}}
\begin{figure}[h]
  \centering
    \includegraphics[width=\linewidth,trim={0 0 0 0},clip=true]
    {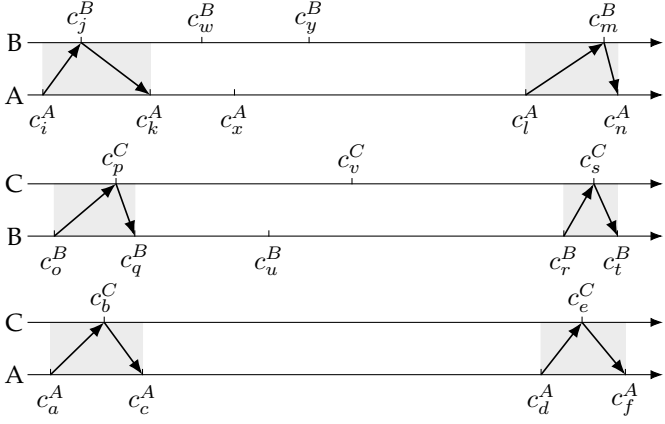}
    \caption[Inter-node duration measurement scheme]
    {%
      Translation examples using \ac{TSC} values from different \ac{MinRTT}
      experiments between node pairs~$\left(A, B\right)$, $\left(B, C\right)$,
      and $\left(A, C\right)$.
    }%
    \label{fig:cp_examples}
\end{figure}

Measuring a \ac{TSC} duration requires different strategies by identifying the
nodes where the start and end events occurred. We distinguish four scenarios:
1) the duration started and ended on the reference node -- the trivial case; 2) the duration
started and ended on the same non-reference node; 3) the duration started and
ended on two nodes including the reference node; 4) the duration started and
ended on two non-reference nodes. While the first case does not incur any
error, the last case incurs the largest. We devised the \ac{CP} method to
exploit each case. For Scenario~3, we compare our method and the baseline.
We explain each scenario using the examples in Fig.~\ref{fig:cp_examples}.
\subsubsection{Trivial case: duration on the reference node}
This case considers a duration consisting of the start and end events that
occurred on the reference node. It does not require translation and thus
incurs no error; it is the same as measuring time durations on
a computer with PAPI.

\subsubsection{Scenario~1: duration on a non-reference node}
The \ac{TSC} duration's start and end events occur on the same node; \CP{1}
leverages it to reduce the size of the error substantially. It translates
Duration~\DurSym{w}{y}{B} into a duration on the reference node~$A$.
\begin{IEEEeqnarray*}{rCl}
  \durSym{w}{y}{B} &=& \tk{y}{B} - \tk{w}{B}
  \\
  \durSym[g]{w}{y}{A} &=& \durSym{w}{y}{B} * \ratioSym{A}{B}
\end{IEEEeqnarray*}
where
\begin{IEEEeqnarray*}{rCl}
  \ratioSym{A}{B} &=& \rFracII{j}{m}{A}{B} = \rFracSym{j}{m}{A}{B}
\end{IEEEeqnarray*}
Each \ac{MinRTT} round trip measures \Tk{j}{A} and \Tk{m}{A}. Refer to
Fig.~\ref{fig:method_cp_ratio} and Section~\ref{sec:refTSC} for more details.
We evaluate the error using Eq.~\eqref{eq:err_prop_rule} and
Eq.~\eqref{eq:chooseMax}, and the resulting error \DurSymErr[b]{w}{y}{A} is as
follows:
\begin{align*}
  \durSymErr[b]{w}{y}{A} = \frac{2\durSym{w}{y}{B}}{\durSym{j}{m}{B}}\var{e}
\end{align*}
\DurSymErr[b]{w}{y}{A} is smaller than \Var[b]{e} because
$2\durSym{w}{y}{B}<\durSym{j}{m}{B}$, where \DurSym{w}{y}{B} is a duration
within \DurSym{j}{m}{B}, which marks the beginning and end of the experiment,
that is substantially long.

\subsubsection{Scenario~2: duration on two nodes, including the
reference node}
Let node~$A$ in Fig.~\ref{fig:cp_examples}  be the reference node; \CP{2}
measures the duration between \Tk{x}{A} and \Tk{y}{B}. This translation incurs
the same error size as the translation explained in
Section~\ref{sec:trans_val}.
\begin{align*}
  \durSymErr[b]{x}{y}{A} = \var{e}
\end{align*}

\subsubsection{Scenario~3: duration on two non-reference nodes}\label{sec:CP3}
Let node~$A$ be the reference node, and consider a \ac{TSC} duration between
\Tk{u}{B} and \Tk{v}{C} in Fig.~\ref{fig:cp_examples}.  It involves invoking
two scenarios, \CP{2} and \CP{1}.  First, perform \CP{2} between nodes~$B$ and
$C$; node~$B$ is the tentative reference node. Then, perform \CP{1} to
translate the duration from node~$B$ to node~$A$.  We begin the evaluation
between nodes $B$ and $C$: let \RatioSym[b]{B}{C} be the \ac{TSC} ratio between
nodes~$B$ and $C$; then use \CP{2}, which results in \DurSym{u}{v}{B}.
\begin{align*}
  \tk[g]{v}{B} &= \tk[g]{p}{B} + \ratioSym{B}{C}\durSym{p}{v}{C}
  \\
  \durSym[g]{u}{v}{B} &= \tk[g]{v}{B} - \tk{u}{B}
\end{align*}
Then use \CP{1} to translate \DurSym{u}{v}{B} into \DurSym{u}{v}{A}.
\begin{align*}
  \durSym[g]{u}{v}{A} &= \ratioSym{A}{B}\durSym[g]{u}{v}{B}
\end{align*}
We determine the error with Eq.~\eqref{eq:err_prop_rule} and
Eq.~\eqref{eq:chooseMax}.
\begin{IEEEeqnarray}{rCl}\label{eq:C4ErrTotal}
  \durSymErr[b]{u}{v}{A}
  &=&
  \frac{2\durSym{u}{\hat{v}}{B}}{\durSym{j}{m}{B}}
  \var{e}
  +
  \ratioSym[b]{A}{B}
  \var{e}
\end{IEEEeqnarray}
We know that $\durSym{u}{\hat{v}}{B} \ll \durSym{j}{m}{B}$, where we set the
time duration between the start and end of a \ac{MinRTT} arbitrarily long;
therefore, the following holds.
\begin{IEEEeqnarray}{rCl}
  0 < \frac{2 \durSym{u}{\hat{v}}{B}}{\durSym{j}{m}{B}} \ll 1
  \nonumber\\
  0 < \frac{2 \durSym{u}{\hat{v}}{B}}{\durSym{j}{m}{B}}\var{e} \ll
  \var{e}\label{eq:C4ErrPart1}
\end{IEEEeqnarray}
Next, the following equation explains that the ratio \RatioSym[b]{A}{B}
requires reading from \acp{TSC} on nodes~$A$ and $B$ at the same \ac{TSC}
instances, $j$, and $m$.
\begin{IEEEeqnarray*}{C}
  \ratioSym[b]{A}{B}
  =
  \frac{\durSym{\hat{j}}{\hat{m}}{A}}{\durSym{j}{m}{B}}
\end{IEEEeqnarray*}
In our cloud configuration, the participating nodes~A and B are hypervised
machines~(HVMs), meaning that their \acp{TSC} run at a different but relatively
similar speed. While exploiting the difference is the key in the \ac{CP}
inter-node time duration measurement methodology, acknowledging their
similarity allows us to evaluate the error size; therefore, the following
comparisons hold.
\begin{IEEEeqnarray*}{rCl}
   &0 \ll \frac{\durSym{\hat{j}}{\hat{m}}{A}}{\durSym{j}{m}{B}} \leq 1
  ,&\quad\text{or}
   \\
   &1 \leq \frac{\durSym{\hat{j}}{\hat{m}}{A}}{\durSym{j}{m}{B}} \ll 2
   &
\end{IEEEeqnarray*}
It follows that the size of error incurred by
$\left(\durSym{\hat{j}}{\hat{m}}{A}/\durSym{j}{m}{B}\right)\var{e}$ is greater
than 0, close to \Var{e}, but smaller than $2\var{e}$.
\begin{IEEEeqnarray}{rCl}
   0 \ll \frac{\durSym{\hat{j}}{\hat{m}}{A}}{\durSym{j}{m}{B}} \var{e} \leq \var{e}
  ,\quad\text{or}\quad
   \var{e} \leq \frac{\durSym{\hat{j}}{\hat{m}}{A}}{\durSym{j}{m}{B}} \var{e} \ll
   2\var{e}\label{eq:C4ErrPart2}
\end{IEEEeqnarray}
Lastly, we can decide the total size of error incurred by \CP{3}.
We evaluated the error size Eq.~\eqref{eq:C4ErrTotal} in two parts:
Eq.~\eqref{eq:C4ErrPart1} and Eq.~\eqref{eq:C4ErrPart2}. We add up the
error beginning with Eq.~\eqref{eq:C4ErrPart2}.
\begin{IEEEeqnarray*}{rCl}
   0 \ll \frac{\durSym{\hat{j}}{\hat{m}}{A}}{\durSym{j}{m}{B}} \var{e} \leq \var{e}
  ,\quad\text{or}\quad
   \var{e} \leq \frac{\durSym{\hat{j}}{\hat{m}}{A}}{\durSym{j}{m}{B}} \var{e} \ll
   2\var{e}
\end{IEEEeqnarray*}
We know that the size of error incurred by Eq.~\eqref{eq:C4ErrPart1} is
significantly smaller than \Var{e}.
\begin{IEEEeqnarray*}{rCl}
  0 &\ll& \left(\frac{2 \durSym{u}{\hat{v}}{B}}{\durSym{j}{m}{B}} +
  \frac{\durSym{\hat{j}}{\hat{m}}{A}}{\durSym{j}{m}{B}}\right) \var{e} \leq \var{e}
  ,\quad\text{or}\quad
  \\
  \var{e} &\leq& \left(\frac{2 \durSym{u}{\hat{v}}{B}}{\durSym{j}{m}{B}} +
    \frac{\durSym{\hat{j}}{\hat{m}}{A}}{\durSym{j}{m}{B}}\right)\var{e} \ll
   2\var{e}
\end{IEEEeqnarray*}
Therefore
\begin{IEEEeqnarray*}{rCl}
  0 \ll \durSymErr[b]{u}{v}{A} \leq \var{e}
  ,\quad\text{or}\quad
  \var{e} \leq \durSymErr[b]{u}{v}{A} \ll
   2\var{e}
\end{IEEEeqnarray*}

\subsubsection{Baseline: duration on two non-reference nodes}
We take the same example used in Section~\ref{sec:CP3} but demonstrate a
relatively trivial approach, where \ac{TSC} events are translated individually
for further arithmetic operations.
The overall process is twofold.  First, we translate \Tk{u}{B} and \Tk{v}{C} as
follows:
\begin{IEEEeqnarray*}{rCl}
  \tk[g]{u}{A}
  &=& \tk[g]{j}{A} + \left(\tk{u}{B}-\tk{j}{B}\right)\ratioSym{A}{B}
  \\
  \tk[g]{v}{A}
  &=& \tk[g]{b}{A} + \left(\tk{v}{C}-\tk{b}{C}\right)\ratioSym{A}{C}
\end{IEEEeqnarray*}
where,
\begin{IEEEeqnarray*}{rCl}
  \ratioSym{A}{B}&=&\rFracSym{j}{m}{A}{B}\text{, }\quad\text{and}
  \\
  \ratioSym{A}{C}&=&\rFracSym{b}{e}{A}{C}
\end{IEEEeqnarray*}
Second, we measure the duration using the translated \ac{TSC} values:
\begin{IEEEeqnarray}{C}\label{eq:baseline}
  \durSym[g]{u}{v}{A} = \tk[g]{v}{A} - \tk[g]{u}{A}
\end{IEEEeqnarray}
Eq.~\ref{eq:baseline} allows us to evaluate the overall error
\DurSymErr[b]{u}{v}{A}.  We determine the error with
Eq.~\eqref{eq:err_prop_rule} and Eq.~\eqref{eq:chooseMax}.  The evaluated error
is as follows.
\begin{IEEEeqnarray*}{C}
  \durSymErr[b]{u}{v}{A} = 2 \var{e}
\end{IEEEeqnarray*}

\subsubsection{Error size comparison}
We evaluated the error size of the trivial case, three scenarios \CP{1},
\CP{2}, \CP{3}, and the baseline in the preceding sections.  The following
table shows the overall comparison between them.
{\setlength\tabcolsep{2pt}
\renewcommand{\arraystretch}{1.3}
\begin{table}[h]
\caption[Error size comparison of the CP methods]
{%
  The baseline incurs the largest error while the error size of \CP{3}, \CP{2},
  and \CP{1} follows in order.
}%
\begin{center}
\begin{tabular}{ c r l c }
\toprule
  \multirow{2}{*}{\ac{CP} method}
  &
  \multicolumn{2}{c}{Size of error}
  &
  \multirow{2}{*}{Range of error \Var{e_n}}
  \\
  &
  \multicolumn{2}{c}{(relative to the max error \Var{e})}
  &
\\
\midrule
  Trivial Case & 
               & -- 
               & $ {\text{--}} $
               \\
  Scenario~1   & $ \durSymErr[b]{w}{y}{A} = $ 
               & $ { 2 \left( \durSym{w}{y}{B} / \durSym{j}{m}{B} \right) } $
               & $ {0 < \var{e_1} < \var{e}} $
               \\
  Scenario~2   & $ \durSymErr[b]{x}{y}{A} = $ 
               & \num{1}
               & $ {\var{e_2} = \var{e}} $
               \\
  \multirow{2}{*}{Scenario~3}
               & $ \durSymErr[b]{u}{v}{A} = $ 
               & $ { 2\left(\durSym{u}{\hat{v}}{B}/\durSym{j}{m}{B}\right) } $
               & $ { 0 \ll \var{e_3} \leq \var{e} \text{ or } } $
               \\
               &
               & $ { + \durSym{\hat{j}}{\hat{m}}{A}/\durSym{j}{m}{B} } $
               & $ { \var{e} \leq \var{e_3} \ll 2\var{e} } $
               \\
  Baseline     & $ \durSymErr[b]{u}{v}{A} = $ 
               & \num{2}
               & $ {\var{e_\mathit{base}} = 2\var{e}} $
               \\
\bottomrule
\end{tabular}
\end{center}
\label{tab:cp_errComp}
\end{table}
}

Table~\ref{tab:cp_errComp} shows that our translation strategy with each
scenario successfully prevents translatation errors from bloating
unnecessarily. Scenario~1 allows translating cycle-level \ac{TSC} durations,
such as computation time on a non-reference node, into the reference \ac{TSC}
cycles.  Unlike the baseline that uses Scenario~2 twice, Scenario~3 combines
Scenario~2 and Scenario~1, which minimizes the incurred error.

\subsection{Timestamp Counter~(TSC)\label{sec:cp_tsc}}

\emph{TSC requirements}.
The \ac{CP} inter-node measurement method approximates the time duration of a
tuple based on \ac{TSC} values and \ac{TSC} frequency retrieved from each cloud
node.  \Ac{TSC} measurement requires the CPU to feature \emph{nonstop\_tsc} and
\emph{constant\_tsc} (a combination of the two is equal to
\emph{invariant\_tsc}). CPUs with the nonstop\_tsc feature guarantee counting
increasing its \ac{TSC} even when the CPU's C-state increases over C2. The
constant\_tsc feature guarantee counting the \ac{TSC} at a constant rate even
when CPU frequency changes. Both features are mandatory and generally available
on modern Intel x86 CPUs.  However, \ac{TSC} may not increase in a lock step on
a system with two CPU sockets or more.

\emph{TSC wrapper function}.
We devised a new function, \emph{clock\_gettime\_tsc}, which calls the
kernel-provided \emph{clock\_gettime} and the \emph{rdtscp} instruction.  For
every tuple that invokes a \acs{bIDh}, the \ac{TSC} value and the
local clock (\emph{CLOCK\_MONOTONIC\_RAW})'s timestamp will be logged for
measuring the duration. (See Section~\ref{sec:handlers} for more details on the
\acs{bIDh}.)

\section{CP Data Generator\label{s:dg}}

\begin{figure*}[t]
  \centering
  \includegraphics[width=\linewidth,trim={0 0 0 0},clip=true]
  {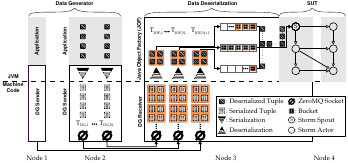}
  \caption[Data deserialization parallelization in the Java object factory]
  {%
    Data deserialization parallelization by the \acf{JOF} across JVM and
    machine code where DG receivers receive serialized tuples emitted by DG
    senders over per-thread ZeroMQ connections.
  }%
  \label{fig:DGOverview}
\end{figure*}

Profiling an \ac{SPE} requires a data generator that can produce data streams
consistently and at a high emission rate.
Existing solutions fail in one or both of these requirements.
\subsection{Kafka Shortcomings}
Apache Kafka~\cite{Kafka2011} is a framework for distributed data generation,
and it provides producers, consumers, and a cluster of brokers. A Kafka
cluster accepts tuple streams generated by one or more Kafka producers; then,
Kafka consumers subscribed to the cluster receive the tuples.
Apache Kafka meets the following requirements. First, it provides a high-level
abstraction of the streaming architecture. Second, it decouples data generation
(Kafka producers) and data consumption (Kafka consumers).  Third, it provides a
common pattern to the industry~\cite{KafkaUsage, ApacheSamza}.
However, Kafka's high-level abstraction comes at the cost of performance.
Prior work~\cite{Karimov18, YahooBottleneck, YahooBenchmark2016,
perera2016reproducible} evaluated Apache Kafka as a data
generator and reported that Kafka became the throughput bottleneck in those
experiments.
Chintapalli~\textit{et al.}~\cite{YahooBenchmark2016} employed multiple Kafka
producers to create the required emission rate because a single Kafka producer
failed to emit more than \SI{17}{\kilo\tuple\per\second}.
Perera~\textit{et al.}~\cite{perera2016reproducible} showed that a single Kafka
broker deployed in the \ac{YhSB} increases the event-time latency by
\SI{30}{\second} when increasing the emission rate higher than
\SI{4}{\kilo\tuple\per\second}.
Hesse~\textit{et al.}~\cite{Kafka:eval} evaluated the performance of Kafka with
their in-house data generators~\cite{Kafka:eval:repo} and determined that
Kafka's throughput is capped at \SI{420}{\kilo\tuple\per\second} when receiving
tuples from two data generators.
Researchers have abandoned Kafka in favor of their data generator design.
Grier~\cite{YahooBottleneck} abandoned Kafka and increased the emission rate by
implementing the data generation facility within the \ac{SPE}.
Karimov~\textit{et al.}~\cite{Karimov18} abandoned Kafka and designed their
\acl{DG}.
\subsection{Consistently High Throughput is Difficult to Achieve}
To produce a data stream at a high and consistent emission rate is a
challenging task with cloud nodes.
Chintapalli~\textit{et al.}~\cite{YahooBenchmark2016} reported that individual
Kafka producers began to fall behind at around \SI{17}{\kilo\tuple\per\second}.
Lu~\textit{et al.}~\cite{StreamBench} point out data generation rate is
different from the data ingestion rate.  Their streaming system, which employs
a messaging system such as Kafka, requires the data generation rate to be
faster than the data consumption rate to keep the \ac{SPE}'s ingestion rate.
Hesse~\textit{et al.}~\cite{Kafka:eval} implemented a data
generator~\cite{Kafka:eval:repo} as a Kafka producer to measure the attainable
ingestion rate of the Kafka cluster. The data generator emits data at a
configurable rate.
To evaluate their data generator without Apache Kafka, we employed the ZeroMQ
throughput test~\cite{ZmqPerfTest}. Instead of using the Kafka producer APIs,
we replaced them with the ZeroMQ \emph{PUSH and PULL}
sockets~\cite{ZmqSocketApi}. We employed two cloud nodes running one data
generator and one JeroMQ~\cite{JeroMQ} client each. The test ran for
\SI{90}{\second}, and their data generator achieved
\SI{400}{\kilo\tuple\per\second} with tuple loss.  We found that the data
generator stops emitting periodically (up to \SI{1.57}{\second}) due to GC
interruptions.  To determine the achievable throughput without tuple loss, we
iteratively lowered the emission rate to \SI{16.67}{\tuple\per\second},
where the data generator could sustain the throughput without tuple loss.
Apache JMeter~\cite{JMeterWeb} is a data generation framework for load testing
various web applications. JMeter's workload model simulates multiple users
(\textit{i.e.}, multiple HTTP sockets) and request-and-reply sequences by each
user. Its highest throughput is \SI{70}{\kilo\request\per\minute} with
\numrange{165}{185} users~\cite{vanHoorn2007thesis}.
Kallas~\textit{et al.}~\cite{DiffStream} proposed a framework for validating
the correctness of processed tuples. They use JUnit-QuickCheck~\cite{JUQCheck}
as a data generator, and they reported a maximum throughput of
\SI{40.082}{\kilo\tuple\per\second} before exhibiting fluctuation. We measured
that their data generator could achieve \SI{326}{\kilo\tuple\per\second} when
emitting tuples to a JeroMQ client.
Karimov~\textit{et al.}~\cite{Karimov18} implemented a distributed data
generator with a configurable data generation rate, and their data generation
rate is faster than the \ac{SUT}'s data ingestion rate. They used the following
techniques to even out the data generation and ingestion rates.  First, they
chose to add a queue between a data generator and a source operator to keep up
with the data generation rate.  Second, they put each pair of data generators
and queues on the same machine to avoid network overhead.  It has two
advantages and a drawback. First, the \ac{SUT} does not lose any tuples because
the queues keep them until the \ac{SUT} acquires them.  Second, the queues
ensure a consistent emission rate by their \acp{DG}.  On the other hand, their
design requires inspection of both the throughput and the event-time latency to
determine the \ac{SUT}'s sustainable throughput. If the \ac{SUT} does not
exhibit prolonged backpressure (\textit{i.e.}, the event-time latency does not
increase continuously), the currently configured generation rate becomes the
sustainable throughput. They need a way to measure lags-per-second. An
ongoing work~\cite{LagTrendMetric} attempts to provide quantification using
lags-per-second.
Fig.~\ref{fig:max_throughput} in Section~\ref{ssec:dg} describes our new
algorithm to evaluate the maximum throughput of the \ac{SUT} correctly.

\subsection{Throughput-controlled Emission Loop \label{ssec:dg}}

\begin{figure}[htb]
  \subfloat[Example][An iterative method to run both DG Sender and DG Receiver
  until the maximum throughput of \ac{SUT} is determined.]%
  {\label{fig:eval_max}%
\begin{lstlisting}[style=PseudoStyle,escapechar=|]
// Supervising all experiments
throughput evaluate_max_throughput(duration d) {|\label{max:begin}|
  for throughput t in range(1M .. 5M, step 1K) {
    nr_emit = DG_Sender(t, d);
    nr_recv = DG_Receiver();
    if (nr_emit != nr_recv) break;
    max_t = t;
  }
  return max_t;
}|\label{max:end}|
\end{lstlisting}%
  }%

  \subfloat[Example2][A timekeeper algorithm for each tuple emission in a tight loop.]%
  {\label{fig:emit_loop}%
\begin{lstlisting}[style=PseudoStyle,escapechar=|,]
// In each DG Sender
emission_loop(t, d) {|\label{loop:begin}|
  start  = now();|\label{var:loop:start}|
  end    = start + d;|\label{var:loop:end}|
  budget = NANO_IN_SECOND / t;
  assert (budget >= ZMQ_Overhead);|\label{loop:assert}|
  t_next = start;
  while (true) {
    t_next += budget;
    socket.send(tup); nr_emit++;
    while ( t_next > now() ) ;|\label{loop:query}|
    if (now() >= end) { break; }
  }
  return nr_emit;
}|\label{loop:end}|
\end{lstlisting}%
  }%
  \caption[Throughput-controlled emission of tuples]
  {%
    The iterative algorithm~(a) runs the emission loop~(b) in each DG Sender
    until the maximum throughput is determined.
  }%
  \label{fig:max_throughput}
\end{figure}

To address the shortcomings of the existing \ac{DG} design, we propose a
measurement methodology (as in Fig.~\ref{fig:max_throughput}) to correctly
evaluate the maximum throughput of the \ac{SUT} across multiple cloud nodes.

\emph{evaluate\_max\_throughput} (Fig.~\ref{fig:eval_max}).
For all experimental runs, we set a duration of $d$ for the current run.
\Ac{SPE} runs infinitely in production nodes; however, our measurement
methodology incorporates duration for bounding the experiment's \emph{start}
and \emph{end} (lines~\ref{var:loop:start} and \ref{var:loop:end}). We increase
$t$ iteratively for each run until a \emph{tuple drop} occurs (\textit{i.e.},
the number of tuples generated by the DG sender must equal the number of tuples
ingested by the \ac{SUT} unless a tuple drop occurred).  A tuple drop indicates
that the last successful throughput is the maximum throughput of the \ac{SUT}.

\emph{emission\_loop} (Fig.~\ref{fig:emit_loop}).
It guarantees to send each tuple within a unit time $\mathit{budget}$ by
querying the current time \emph{now()} while it is within the time window
$[t_{\mathit{next}}-\mathit{budget}, t_{\mathit{next}}]$
(line~\ref{loop:query}). It ensures the emission loop will send all tuples
before the designated time \emph{end\_loop}.

\emph{Overhead of a ZeroMQ send call}.
We ensured that a $\textit{budget}$ is equal to or bigger than the overhead of
ZeroMQ's send operation \emph{ZMQ\_Overhead} (line~\ref{loop:assert}). We
measured that the single-call overhead of ZeroMQ's send operation is
\SI{1.90}{\micro\second}, \textit{i.e.}, one emission thread will not run
faster than \SI{0.53}{\mega\tuple\per\second}.
It is \SI{141.06}{\mebi\byte\per\second} using a \ac{YhSB} tuple, and a total
of nine ZeroMQ threads can emit at \SI{4.41}{\mega\tuple\per\second}
(\SI{9.79}{\giga\bit\per\second}) at maximum.

\subsection{Tuple Processing with the Java Object Factory\label{ssec:tuple}}
In Fig.~\ref{fig:DGOverview}, DG Sender runs on Nodes~2 and 3 attached to a
user application. Assume that on Node~2 in Fig.~\ref{fig:DGOverview}, a user
application running on JVM creates an application-specific tuple as a Java
object (\textit{E.g.}, \ac{YhSB} creates advertisement information events
exposed to a user). We designed DG Sender as \ac{JNI} machine code built with
the user application. DG Sender requires the application to pass the tuple data
as a byte array using Kryo~\cite{KryoGitHub} for Java object serialization.
(Apache Spark~\cite{SparkTuning} and Apache Storm use Kryo.) The application
hands over the resulting byte array to thread~$T_{\mathit{DG},1}$ via the
\ac{JNI} interface provided by DG Sender. DG Sender prepares all tuples
serialized ahead of the experiment to avoid Java object serialization overhead
from the critical path in the emission loop. In each emission thread, the
ZeroMQ socket sends the serialized tuple to the corresponding ZeroMQ socket in
a \ac{JOF} thread~$T_{\mathit{JOF},1}$ on Node~3.  \Ac{JOF} is our Java object
deserialization library built against the \ac{SUT} as part of \ac{CP}. Each
\ac{JOF} thread runs a deserialization loop that pulls out buckets of
serialized tuples and pushes each bucket to an SPSC queue after de-serializing
all tuples in the bucket.  Here, we use Kryo for deserialization, and we take
all overhead of deserialization which affects the resulting throughput.  We
deploy one JCTools~\cite{JCTools} SPSC queue in each \ac{JOF} thread. A Storm
Spout in the \ac{SUT} iteratively drains the queues in the \emph{nextTuple()}
method, fetching one bucket from a queue and ingesting a tuple at a time.

Using \ac{JOF}, we measured that the \ac{SUT} can consistently ingest tuples at
\SI{4.68}{\mega\tuple\per\second} without exhibiting a tuple drop.
It is five times faster than running the same \ac{SUT} without \ac{JOF}
(\SI{0.9}{\mega\tuple\per\second}).
We ran the experiment on seven \ac{GCE} nodes -- three for ZooKeeper, one for
Redis , one for Storm Nimbus, one for DG Sender, and one for \ac{SUT}.  We
implemented a Storm sink application as the \ac{SUT}. It receives tuples in the
Spout and drains them in the subsequent actor.

\section{CP Profiling Framework\label{sec:cp_lib}}

\Ac{CP} profiling framework is a native \ac{C++}11 library that implements
\ac{TSC} inter-node measurement method with cloud profiling components.

The latency of event processing times of a streaming engine provides insights
into its operational efficiency over time.  Because data arrives at high
density and velocity and loses value over time, event processing latency is
crucial for the perfromance engineering of an \ac{SPE}.

We tag each event with a tuple~ID to trace them and their
processing times as they flow through a stream graph topology. We explored
relevant software engineering exercises.
\emph{Retroactive extension}~\cite{TypeClasses} describes a software
engineering requirement that allows extending a program without recompiling
the source code.
\emph{Manifold}~\cite{Manifold} is an ongoing attempt that implements such
software engineering requirements for Java; however, we discovered that its
\emph{Extension} feature only supports adding a static field to an existing
class, whereas we need to inject a unique tuple~ID in each tuple object. Thus,
we added two APIs in \ac{CP} for application developers: \emph{getBytes()} and
\emph{loadBytes()}.
A tuple receives a \SI[number-unit-product=\text{-}]{64}{\bit} tuple~ID using
\emph{getBytes()}, and the function serializes the tuple into a byte array.
\Ac{CP} then preloads the byte array for data generation with
\emph{loadBytes()}.

To avoid perturbing the Java code running on the streaming engine, we designed
\ac{CP} native library to conduct all high-resolution time measurements in
machine code. As depicted in Fig.~\ref{fig:cpjava}, our approach is minimally
invasive.  The \ac{SUT} only needs to open \emph{channels}, which to the JVM,
are represented as values of primitive type \emph{long}.  (We defer the
discussion of the channel's handler argument to Section~\ref{sec:handlers}.)
After an invocation of the function \emph{openChannel()}
(line~\ref{cpjava:open}), \emph{logTS()} (line~\ref{cpjava:log}) allows logging
a $\langle$timestamp, tuple~ID$\rangle$ pair. The \ac{CP} framework stores such
log information in each channel's underlying log file. Log files are collected
post-mortem for retrieval of profiling information.

Assigning each event a unique tuple~ID and inserting instrumentation similar to
Fig.~\ref{fig:cpjava} make it possible to trace a tuple through the
stream-graph topology at nanosecond resolution.  Particular points of interest
and instrumentation in the application code are the start and end of an actor's
work function and calls to receive and post tuples.

\subsection{Handlers and Trace- and Sample-based Profiling}\label{sec:handlers}

\begin{figure}[tbh]
\begin{lstlisting}[style=JavaStyle,escapechar=|,]
import cp.*;
// ...
System.loadLibrary("Cloudprofiler");|\label{cpjava:load}|
// ...
long ch0 = Cloudprofiler.openChannel(|\label{cpjava:open}|
    "PC handler #1",|\label{cpjava:name}|
    log_format.ASCII,
    handler_type.NET_CONF);|\label{cpjava:handler}|
// ...
Cloudprofiler.logTS(ch0, tuple_id);|\label{cpjava:log}|
// ...
Cloudprofiler.closeChannel(ch0);
\end{lstlisting}
    \caption
    {%
      Deployment of the \ac{CP} in a Java streaming application
    }%
    \label{fig:cpjava}
\end{figure}
\begin{figure}[tbh]
\begin{lstlisting}[style=CStyle,escapechar=|,]
struct closure {
  void (*handler_f)(void *closure,|\label{cl:handler}|
                    int64_t tuple);
  int  (*parm_handler_f)(void *closure,|\label{cl:param}|
                         int32_t arg_num,
                         int64_t arg_val);
  int32_t       ch_nr;        // channel number
  std::ofstream ch_logf;      // log-file
  std::string  *ch_logf_name; // log-file name
};
\end{lstlisting}
  \caption
  {%
    Closure representation of a handler in the \ac{CP} library
  }%
  \label{fig:closure}
\end{figure}

The \ac{C++}11 native APIs are exported to \ac{JNI} using the SWIG-Java
library~\cite{SWIGPaper}.  Thus, we can instrument both \ac{C++}11 and Java
programs with our \ac{CP} library.  We demonstrate the use of the library in
Fig.~\ref{fig:cpjava}. The \ac{CP} native library loads once per JVM instance
(line~\ref{cpjava:load}).  Opening a profiling channel requires a channel name,
log format, and handler type (lines~5--8).  For the current channel \emph{ch0},
the library logs a $\langle$\ac{TSC}~value, tuple~ID$\rangle$ pair, effectively
distinguishing each tuple (line~\ref{cpjava:log}).  The timestamp comes from
the \emph{rdtscp()} instruction.  Closing a channel causes buffered logging
data (if any) to be flushed to a disk.  As depicted in Fig.~\ref{fig:closure},
a closure~\cite{Closures} represents a profiling channel, which contains a
pointer to the channel's handler function (line~\ref{cl:handler}) and the
entire referencing environment (data) of the handler.  (Complex handlers,
\textit{e.g.}, an FIR filter, can extend the closure type by adding the
required book-keeping information.) A handler function receives its closure as
a type-erased argument~\cite{IntenPolym} to avoid the overhead of a dynamic
(dispatching) method call.  Internally, the channel~ID of type \emph{long} from
Fig.~\ref{fig:cpjava} is a pointer to the channel's closure. Procedure
\emph{logTS()} calls the closure's handler with the closure as its argument.
Handler functions are type-specific and thus able to regain the closure type
through a \ac{C++} static cast~\cite{c++11}.  We optimized the handler
dispatching mechanism for logging events in streaming workloads at
nanosecond-level accuracy.  The \emph{parm\_handler\_f} function pointer
(line~\ref{cl:param}) lets handlers incorporate a function that parameterizes
the handler.  We provide several types of handlers to conduct trace- and
event-based sampling. A handler associates with a channel opened (see
Fig.~\ref{fig:cpjava}, line~\ref{cpjava:handler}).

\emph{\Ac{IDh}}.
This handler writes the $\langle$ \ac{TSC}~value, tuple~ID$\rangle$ pair
verbatim to the channel's underlying log file.

\begin{figure*}[t]
  \centering
  \includegraphics[width=\linewidth,trim={0 0 0 0},clip=true]
  {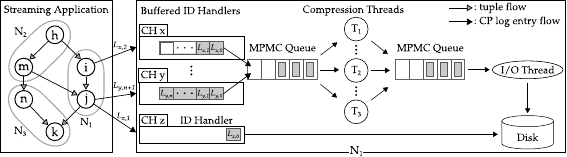}
  \caption[CP run-time environment]
  {%
    \ac{CP} run-time environment: comparison between the \acs{bIDh} and the
    \ac{IDh}; the \ac{IDh} directly writes to the disk, whereas the
    \acs{bIDh} buffers and compresses data for higher disk I/O bandwidth.
  }%
  \label{fig:id_vs_buffer_id}
\end{figure*}

\emph{\Ac{bIDh}}.
It buffers the $\langle$TSC~value, tuple~ID$\rangle$ data, performs in-memory
compression, and writes them to the channel's log file.  As shown in
Fig.~\ref{fig:id_vs_buffer_id}, the handler initializes with multiple blocks,
and each block can store \SI{1}{\million} tuples (a compile-time constant).
Instead of writing each logged tuple directly into the channel's log file, it
buffers tuples into the blocks. We employ the non-blocking Michael-Scott
queue~\cite{MSQueue} from the Boost \ac{C++} libraries~\cite{BoostWeb} for
moving blocks.  A block is enqueued into the boost MPMC queue when it becomes
full.  When a block is full, it is enqueued into the boost MPMC queue.  We
employ multiple compression threads to dequeue and compress blocks and enqueue
them into the I/O queue (another Boost MPMC queue).  The I/O thread
dequeues the compressed blocks and writes them into the corresponding log file
with its channel.  For termination, \ac{CP} registers a callback function that
activates with a \emph{SIGTERM} signal. Termination begins with closing all
associated channels; it stops ingesting new data and flushes the
processed data left in the system, ensuring no log data is lost.

\emph{\Ac{PCh}}.
It counts the number of tuples read within a period of specified duration. At
the beginning of each period, the periodic counter resets and starts from zero.
When the channel closes, all periodic counts accumulates into the total count.
The \ac{PCh} consumes fewer resources than the \acp{IDh} and determines the
changes in throughput during the benchmark run. We employ a sampling thread
that maintains the start of all periods in a lock step. We propose two versions
of the \ac{PCh}. First, a \emph{\ac{SPCh}} provides optimized performance when
deployed on a dedicated core. It uses a relaxed memory operation for counting
tuples and ensuring the value is globally visible when the sampling thread
accesses the variable. Second, a \emph{\ac{MPCh}} uses acquire-and-release
memory operations, which gains performance across threads.  We found that the
\ac{MPCh} runs at
\SI[round-mode=places,round-precision=2]{105.040255}{\mega\tuple\per\second}
while the \ac{SPCh} runs at
\SI[round-mode=places,round-precision=2]{305.042118}{\mega\tuple\per\second},
increasing the performance three. The complete result is in
Table~\ref{tab:kieker}.

\emph{Downsample handler}.
This handler logs every $n^\text{th}$ tuple/timestamp to the log file and
discards tuples otherwise to reduce profiling overhead.  A channel acquires the
argument~$n$ via the channel's parameter handler.

\emph{XoY handler}.
It receives two arguments: $x$ and $y$. It logs a $\langle$timestamp,
tuple~ID$\rangle$ pair to the underlying file \textit{iff.}
$$\text{tuple\_ID}\mod y < x$$ For example, assume $x=2$ and $y=1024$. The
handler logs tuples~\numlist[list-final-separator={,
}]{0;1;1024;1025;2048;2049}, and subsequent. This handler allows conducting
sample-based logging tailored to \emph{batches} of tuples.  The advantage of
this tailored logging is that the same batch of tuples will show on all
channels that use the same argument values~$x$ and $y$; thereby, it becomes
possible to compute durations for a given tuple.  (The calculation is
impossible if a tuple were kept on one channel but discarded on the next
channel that this tuple passed through.)

\emph{\Ac{FLh}}.
It writes $\langle$tuple~ID, TSC$\rangle$ pairs of the first and last
tuples processed by the streaming application. The handler avoids the overhead
of logging all tuples and provides the first and last timestamps for
calculating the overall tuple processing time.

\emph{\Ac{ReTh}}.
The handler reproduces the \ac{ReT} measurement of a streaming application. We
described the \ac{ReT} method in Section~\ref{ssec:ret_method}.

\emph{Null handler}.
It discards tuples from being logged. It provides a convenient way to
temporarily disable unused channels via the \ac{cnfsvr} (See below for a
detailed explanation) without rebuilding and deploying the target application
on cloud nodes.

\emph{Online Handler Configuration}.
\Ac{CP} provides online handler configuration via the \ac{cnfsvr}. It

opens a handler channel, it retrieves the handler type associated with the
channel name.

All artifacts, including the native \ac{CP} library and the measurement data,
are available~\cite{CPRepo}.

\subsection{MinRTT Measurement\label{sec:minrtt}}

\Ac{CP} provides server and client applications for the \ac{MinRTT}
measurement: \emph{\ac{TSCma}} and \emph{\ac{TSCsl}}.  We employ a \ac{TSCma}
that conducts the measurement across \acp{TSCsl}. We employ one \ac{TSCsl} on
each node inside the experiment; its task is to retrieve a \ac{TSC} value from
a corresponding \ac{TSCsl} while measuring the \ac{RTT}.  The goal of this
measurement is to achieve the following information.

\emph{\ac{MinRTT}}.
The \ac{MinRTT} of two nodes \Node{A} and \Node{B} determines a temporal order
of \ac{TSC} values measured on the two nodes. A \ac{MinRTT} packet that leaves
and comes back to \Node{A} via \Node{B} generates three \ac{TSC} values --
start and end \ac{TSC} values on \Node{A} and a \ac{TSC} value on \Node{B} that
happened between the start and the end. \Ac{MinRTT} conducts several \acp{RTT}
and selects one with the \acl{MinRTT}.

\emph{\ac{TSC} frequency}.
We use \ac{TSC} frequency when representing our experimental data in seconds
rather than the \ac{TSC} cycles. We conduct all experiments in native \ac{TSC}
cycles and use the \ac{CP} method to translate them to the reference \ac{TSC}
cycles. Only then we use the reference \ac{TSC} frequency to increase
readability. We do not use the kernel-calculated \ac{TSC} frequency but
calculate it ourselves using two \acp{MinRTT} (\MinRTT{1} and \MinRTT{2} in
Fig.~\ref{fig:method_cp_ratio}) to tighten the error bounds of oscillators
aging and prone to thermal changes~\cite{Najafi2021}.

Once the \ac{TSCma} has all \acp{TSCsl} registered, it requests \ac{MinRTT}
measurements from every \ac{TSCsl} pair. The \ac{MinRTT} measurement considers
directions ${A}\to{B}$ and ${B}\to{A}$ because network conditions and routes
may vary.  Upon receiving a \ac{MinRTT} request, each \ac{TSCsl} pair runs
several iterations of \ac{RTT} measurement and send back the \ac{MinRTT} value
to the \ac{TSCma}.  By the end of all \ac{MinRTT} measurements, we obtain
\ac{MinRTT} for every node pair.

\section{Validation\label{s:val}}

\subsection{Validation Setup}

We ran all validations on cloud nodes in \ac{GCE}. We describe setups for each
validation.

\emph{Non-virtualized TSC}.
To validate whether \ac{TSC} is virtualized or not, we employed one node . For
comparison, we used Intel Skylake i-5 6600 machine.

\emph{\acs{ReT} method}.
We employed two~nodes for a \ac{ReT} method validation.  \Ac{ReT}
method requires two nodes: one for \ac{ReT}-Start and the other for
\ac{ReT}-End. Each has eight vCPUs and \SI{16}{\gibi\byte} RAM. We wrote
the validation in \ac{C++}.

\emph{\Ac{CP} method}.
We employed 30~nodes to run the \ac{CP} method with the
\emph{\ac{YhSB}}~\cite{YahooBenchmark2016}.  \Ac{YhSB} is an off-the-shelf
benchmark that provides a baseline environment to execute the company's
in-house streaming application on a single server, and we extended it to suit
our multi-node configuration.  Our cloud resource includes 488~vCPUs and
\SI{648}{\gibi\byte} of RAM; the Storm framework uses over two-thirds of the
vCPU resources.  Each of the 10~Storm worker nodes has 32~vCPUs and
\SI{32}{\gibi\byte} RAM, and one node exlusively runs a Storm Nimbus, which is
responsible for scheduling.  Another 10~nodes run cloud infrastructure
applications, including ZooKeeper, Kafka messaging queue, and Redis in-memory
database. The remaining 10~nodes run data generators that feed data into Kafka
brokers. The benchmark is available online on
GitHub~\cite{YahooBenchmarkGithub}.
We selected Apache Storm~2.0.0 to run the benchmark application.  We used
ZooKeeper~3.5.5, Kafka~2.3.0, Redis~4.0.11, and Storm.  All JVM
processes of the benchmark ran on Oracle Java~8, and all nodes ran Cent OS~7
Linux.
We used GCC~8.3.0 to build \ac{CP}-instrumented cloud applications with
libraries SWIG~2.0.10, ZeroMQ~4.1.4~\cite{zeromq}, and Boost~1.61.0.

\emph{\Ac{CP} compression}.
To validate the bandwidth and scalability of the ZSTD compression in \ac{CP}'s
\ac{bIDh}, we employed a node with 56~vCPUs, \SI{256}{\gibi\byte} of RAM, and a
local scratch SSD.

\emph{\Ac{CP} overhead}.
To validate the performance overhead of \ac{CP}, we employed 10~nodes to run
the \ac{YhSB}.

\subsection{TSC Validation in the Cloud \label{s:val_tsc}}

Invoking \emph{rdtsc} and \emph{rdtscp} should read the underlying CPU's
\ac{TSC} register, and we need to validate that \ac{GCE}'s KVM does not
virtualize \ac{TSC} instruction results. First, we ran rdtsc and rdtscp
instructions iteratively \num{10000} times.  Second, to prevent out-of-order
execution by the CPU, we reran the same test \SI{10}{\billion} times.

Invoking rdtsc on a cloud node took \SI{10}{\nano\second} for both iteration
setups.  Running it on an Intel~i5-6600 took \SI{9}{\nano\second} for
\num{100000} iterations and \SI{6}{\nano\second} for \SI{10}{\billion}
iterations.
We ran the same test for rdtscp, which took \SI{16}{\nano\second} on the cloud
node. On an Intel~i5-6600, one invocation took \SI{14}{\nano\second} for
\num{100000} iterations and \SI{8}{\nano\second} for \SI{10}{\billion}
iterations.
The comparison between a cloud node and the i5-6600 bare metal system shows
that \ac{GCE}'s KVM does not virtualize either rdtsc or rdtscp, and the
instructions are usable for our \ac{CP} method. Invoking rdtscp is slower than
rdtsc because it provides additional information, such as CPU processor and
socket numbers~\cite{intelSysVol2}.

\subsection{CP Compression Bandwidth and Scalability\label{exp:comp}}

\begin{figure}[tbh]
  \centering
  \includegraphics[width=\linewidth,trim={0 0 0 0},clip=true]
  {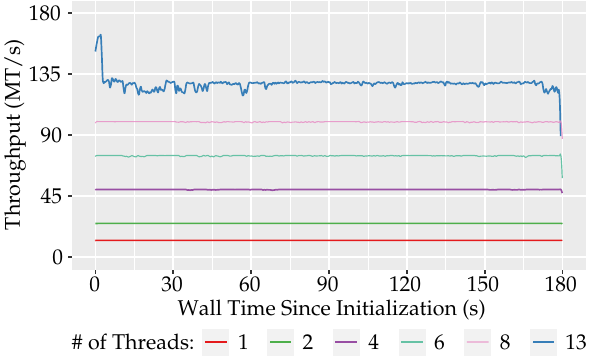}
  \caption[Buffered ID handler throughput, ZSTD codec]
  {%
    Throughput of the \ac{bIDh} with the ZSTD codec. Our benchmark compares the
    performance employing a different number of log threads.
  }%
  \label{fig:buf_throughput1}
\end{figure}

\begin{figure}[tbh]
  \centering
  \includegraphics[width=\linewidth,trim={0 0 0 0},clip=true]
  {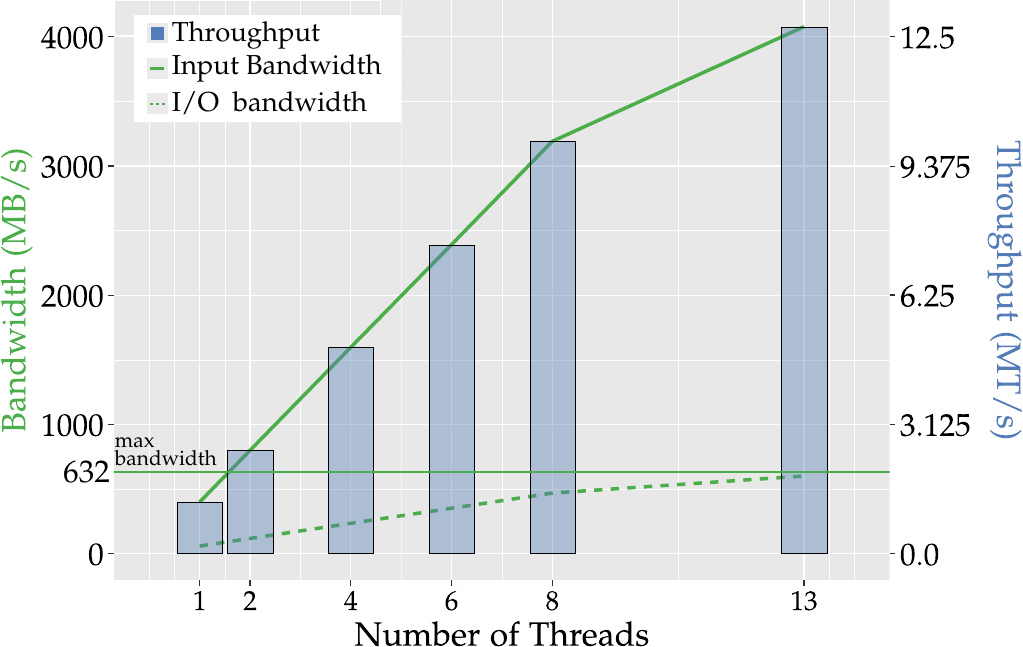}
  \caption[Buffered ID handler scalability, ZSTD codec]
  {%
    Scalability of the \ac{bIDh} with ZSTD codec
  }%
  \label{fig:buf_throughput2}
\end{figure}

Figures~\ref{fig:buf_throughput1} and~\ref{fig:buf_throughput2} show the
scalability of the \emph{\ac{bIDh}}. We use the Squash compression
benchmark~\cite{Squash} to run and compare different compression algorithms.
We chose the \emph{ZSTD} compression algorithm because it shows a high
compression rate and fast compression speed for compressing \ac{CP}'s \ac{IDh}
logs.

We validated \ac{CP}'s logging compression and scalability performance on a
node with a local scratch SSD disk.  Google reported the sustained throughput
limit of \SI{660}{\mega\byte\per\second} and \SI{350}{\mega\byte\per\second}
for read and write operations, respectively.  We also benchmarked the local
scratch SSD disk using the \emph{fio}~tool~\cite{FIO} and measured the maximum
disk bandwidth of \SI{632}{\mega\byte\per\second} for sequential write
operations.  We used the same fio parameters that Google
provided~\cite{BenchmarkSSD}, but we did not set \emph{direct} and
\emph{end\_fsync} to make the behavior of the I/O operations identical to
\ac{CP}'s I/O.

We set each thread to write tuples at \SI{12.5}{\mega\tuple\per\second}
constantly and scale the number of threads until we get the maximum disk
bandwidth.  As shown in Fig.~\ref{fig:buf_throughput1}, we can achieve
scalability up to eight threads, but the test with 13~threads cannot achieve
the required throughput of \SI{162.5}{\mega\tuple\per\second}.

In Fig.~\ref{fig:buf_throughput2}, \Ac{bIDh} achieves
\SI{600.73}{\mega\byte\per\second} with 13~threads, which is close to the
maximum disk bandwidth of \SI{632}{\mega\byte\per\second} that we measured with
fio tool. Note that they are compression performances; the uncompressed data
throughput is \SI{3191}{\mega\byte\per\second}. The ZSTD codec performed the
compression ratio of \num{6.81}.

{\setlength\tabcolsep{3pt}
\renewcommand{\arraystretch}{1.3}
\begin{table}[tbh]
\caption[Single-thread performance of CP handlers]
{%
  Single-thread performance of \ac{CP} handlers in latency and throughput
}%
\begin{center}
\begin{tabular}{ r r r r r r }
\toprule
\multicolumn{2}{r}{}&\multicolumn{2}{c }{\ac{C++}}&\multicolumn{2}{c}{Java}\\
\cmidrule(lr){3-4} \cmidrule(lr){5-6}
\multicolumn{2}{r}{}&Latency&Throughput&Latency&Throughput\\
\multicolumn{2}{c}{100M Iterations}
&\multicolumn{1}{c}{($\text{ns}$)}&\multicolumn{1}{c}{($\text{MT}/\text{s}$)}&\multicolumn{1}{c}{($\text{ns}$)}&\multicolumn{1}{c}{($\text{MT}/\text{s}$)}\\

\midrule
\multirow{9}{*}{\rotatebox[origin=c]{90}{\ac{CP} Handlers}}
 & ID            & $377$ & \fnum{2.652520}   & $472$ & \fnum{2.118644}  \\
 & Buf. ID bin   & $ 63$ & \fnum{15.873016}  & $152$ & \fnum{6.578947}  \\
 & Buf. ID zstd  & $ 73$ & \fnum{13.698630}  & $173$ & \fnum{5.780347}  \\
 & Buf. ID lzo1x & $ 65$ & \fnum{15.384615}  & $141$ & \fnum{7.092199}  \\
 & Downsample    & $175$ & \fnum{5.714286}   & $276$ & \fnum{3.623188}  \\
 & XoY           & $ 13$ & \fnum{72.400649}  & $108$ & \fnum{9.259259}  \\
 & SPC           & $  3$ & \fnum{305.042118} & $ 96$ & \fnum{10.416667} \\
 & MPC           & $  9$ & \fnum{105.040255} & $101$ & \fnum{9.900990}  \\
 & Null          & $  3$ & \fnum{294.727547} & $ 97$ & \fnum{10.309278} \\
\bottomrule
\end{tabular}
\end{center}
\label{tab:cp}
\end{table}
}

As shown in Table~\ref{tab:cp}, we compared the single-thread performance of
various \ac{CP} handlers on a cloud node.  \Acp{PCh} -- \ac{SPCh} and \ac{MPCh}
outperform other handler types because they only log the number of
\emph{logTS()} calls for each unit time. A \ac{bIDh} does the same logging like
an \ac{IDh} does, but it outperforms and logs at up to several million tuples
per second with the help of the underlying compression threads.  The \ac{JNI}
overhead causes all \ac{CP} handlers to exhibit lower throughput on a Java
program than on a \ac{C++} program.

\subsection{ReT Method Validation in the Cloud\label{exp:ret}}

We compared our \ac{CP} measurement method with \ac{NTP} and \ac{ReT} to
validate its accuracy.  To reproduce the \ac{ReT} measurement, we added
\ac{ReT}-Start and \ac{ReT}-End handlers to the \ac{CP} library.  We ran
\ac{ReT}-Start and \ac{ReT}-End programs on two cloud nodes.

\begin{figure}[tbh]
  \centering
  \includegraphics[width=\linewidth,trim={0 0 0 0},clip=true]
  {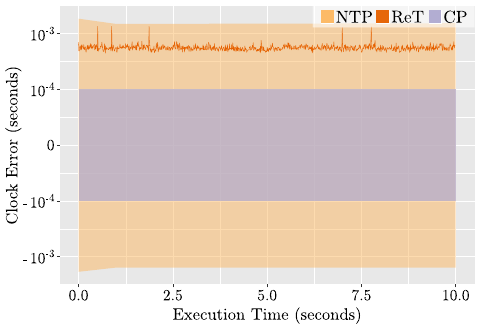}
  \caption[Clock error comparison with a ReT method validation]
  {%
    A comparison of clock errors, including a \ac{ReT} method validation
  }%
  \label{fig:errbnd_cloud_ret}
\end{figure}

Fig.~\ref{fig:errbnd_cloud_ret} depicts that \ac{CP} achieved the lowest and
constant error~bound $\pm\SI{47590}{\nano\second}$.  \Ac{ReT} method resulted
in the average lower~bound of \SI{266.390}{\micro\second} with minimum and
maximum lower~bounds of
\SIlist[list-units=single]{211.646;648.667}{\micro\second}, respectively.
\Ac{NTP}'s clock~error decreases over time during the validation, which marks a
minimum lower~bound of \SI{-889.47}{\micro\second} and a maximum upper~bound of
\SI{710.76}{\micro\second}.  A detailed account of how we determine the
clock~error of the \ac{CP} method on cloud nodes is in Section~\ref{val:yahoo}.

\begin{figure*}[t]
  \centering
  \includegraphics[width=\linewidth,trim={0 0 0 0},clip=true]
  {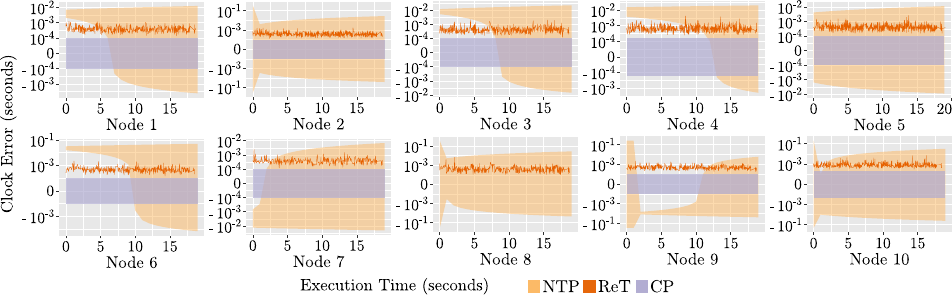}
  \caption[Clock error across worker nodes running the Yahoo streaming benchmark]
  {%
    Comparisons of clock errors on each worker node running the \ac{YhSB}.
    Clock error is lower if: (1) the value is closer to 0, (2) the height of
    the bar is shorter.
  }%
  \label{fig:errbnd_cloud_yahoo}
\end{figure*}

\subsection{Validation with the Yahoo Streaming Benchmark\label{val:yahoo}}

To validate the \ac{CP} method in a production-level environment, we ran the
\ac{YhSB} with \ac{CP} instrumented in the source code and compared its
error~bound to the error~bounds of \ac{NTP} and \ac{ReT} methods.  We ran the
validation on \ac{GCE} nodes with the following design considerations.

\emph{\Acf{MinRTT}}.
\Acp{TSCsl} on all possible unique pair of nodes measure \ac{MinRTT}.  Each
pair exchanges measurement packets 100 times to select \ac{MinRTT} for the
pair. The \ac{TSCma} receives all \ac{MinRTT} results.

\emph{\ac{NTP} clock~error}.
To calculate the error~bound of the clock error at the current second,
\emph{system\_time\_offset}, \emph{root\_delay}, and \emph{root\_dispersion}
values are retrieved by invoking the \emph{chronyc} command every second for
the entire run of each cloud node.

\emph{\ac{ReT} method error~bounds}.
We implemented and incorporated the \ac{ReT} handler in \ac{CP} to run the
\ac{ReT} method. The handler consists of \Ac{ReT}-Start and \ac{ReT}-End
handlers, and they measure the \ac{RTT} of each ReTACK packet.

\emph{\ac{CP} instrumentation}.
We instrumented the \ac{YhSB} with \ac{CP} handlers.  We added a \ac{ReT}-Start
handler to the \emph{DGSpout} actor and a \ac{ReT}-End handler to the benchmark
application's \emph{CampaignProcessor} actor.  DGSpout receives all incoming
tuples coming to the streaming system, and \ac{ReT}-Start handler assigns each
of them with a start time; therefore, it is where all ReTACK packets return to
and receive an end timestamp.  When a tuple arrives in CampaignProcessor class,
the \ac{ReT}-End handler sends back a ReTACK packet to the \ac{ReT}-Start
handler in the DGSpout. The \ac{ReT}-End handler measures the \ac{RTT}, and
both handlers log the start and end timestamps.

In this experiment, the benchmark ran for \SI{19}{\second} after Storm
scheduled the streaming application, mapping it across 10~worker nodes.
The \ac{ReT} method intends to run on a single source actor; therefore, we
configured Storm to allocate one DGSpout. In Fig.~\ref{fig:errbnd_cloud_yahoo},
DGSpout was allocated to Node~8 to receive tuples from the Kafka cluster.

Fig.~\ref{fig:errbnd_cloud_yahoo} shows the clock errors of the compared
methods.  We prepared diagrams for individual nodes because the \ac{TSC} count
and \ac{TSC} increment in each worker node have unique values; likewise, for
two computers with the same CPU, their \ac{TSC} value and \ac{TSC} increment
will not be the same; booting time will make the \ac{TSC} count different, and
CPUs on the two computers will run at a different speed; also, \ac{NTP}
clock~error of each compared nodes is always different because of the network.

Diagrams in Fig.~\ref{fig:errbnd_cloud_yahoo} show that the \ac{NTP}
clock~error is unstable and increases over time on every worker node during the
experiment.  The diagram is also similar to Fig.~\ref{fig:errbnd_cloud_ret}.
The \ac{CP} method retrieves required system information, including each node's
\ac{TSC} count and \ac{TSC} increment during the quiescent time -- before,
after, and in-between network communication bursts; however, \ac{ReT} and
\ac{NTP} method must be running in real-time, during the experiment.  The
error~bound of the benchmark shows a crucial advantage inherent to the \ac{CP}
method.

The \ac{NTP} clock-error bar in nodes~\numlist[list-final-separator={, and
}]{1;3;4;6;9} does not include \num{0} during a specific period of execution
time because the system clock is away from the desired global clock in the
\ac{NTP} hierarchy. The \ac{NTP} protocol monitors the system time offset from
the global clock and applies this information to re-calibrate its system clock.
Global clock information, on the other hand, is transferred in \ac{NTP} packets
sent from peer clients on the network in the subscribed \ac{NTP} hierarchy.
Because an \ac{NTP} packet includes information about the network uses, a less
congested network will contribute less to clock errors.  However, a streaming
application that sends more than \num{100000} data packets per second between
nodes cannot expect a quiescent network.  Because the system wall clock is
consistently re-calibrated by the \ac{NTP} client, our evaluation is to avoid
using the system clock in measuring durations between cloud nodes and on a
stand-alone system where the required measurement accuracy is sub-millisecond.

\Ac{ReT} method, by its design, accumulates half the \ac{RTT} at minimum
when measuring a time duration across multiple nodes. The \ac{ReT} error~bounds
depicted in Figures~\ref{fig:errbnd_cloud_yahoo} and~\ref{fig:errbnd_cloud_ret}
address the glaring overhead.

Conversely, the \ac{CP} method exhibits the smallest and constant error~bound
over execution time. As mentioned, we can retrieve the required \ac{TSC}
values in a quiescent period.

Node~8 of Fig.~\ref{fig:errbnd_cloud_yahoo} does not include \ac{CP}
error~bound because a DGSpout actor resides created in the node.  \Ac{CP}
method is not measurable for a tuple that has ended up in the actor where it
originated. Our implementation of \ac{ReT}-method does not consider this case.
The original work does not mention specific cases but only cases where
\ac{ReT}-Start and \ac{ReT}-End nodes differ.

The result shows that \ac{CP}'s smallest error bound was between
\SIrange{-39722}{39722}{\nano\second}, and its largest error bound was between
\SIrange{-53252}{53252}{\nano\second}.  \Ac{ReT} method showed an average of
\SI{206.337}{\micro\second}, \SI{38.437}{\micro\second}, and
\SI{2612.069}{\micro\second} as the minimum and the maximum, respectively.
\Ac{NTP} method's clock~error showed \SI{-182.089}{\milli\second} and
\SI{182.118}{\milli\second} as its minimum lower~bound and maximum upper~bound,
respectively. (We do not use the results from those nodes where the wall clock
offsets have gone off from \num{0}.) The biggest absolute error-bound size
measured was \SI{364.208}{\milli\second} from the previously selected minimum
lower~bound and maximum upper~bound. This pair was measured on Node~2 at the
first second into the benchmark execution. \Ac{MinRTT} may vary on different
instantiations of the same node. The maximum \ac{MinRTT} we measured is
\SI{226627}{\nano\second}.

\subsection{CP Instrumentation Overhead\label{val:CPOverhead}}

{\setlength\tabcolsep{6pt}
\renewcommand{\arraystretch}{1.15}
\begin{table}[tbh]
  \caption[Throughput comparison with and without the \ac{CP}
  instrumentation]{%
  Throughput comparison with and without the \ac{CP} instrumentation
  }%
  \small
  \centering
  \sisetup{round-mode=places, round-precision=1,
  scientific-notation=fixed, fixed-exponent=3,
  table-omit-exponent}
  \begin{tabular}{ c S S S S c }
  \toprule
  Configurations & \text{Min} & \text{Max} & \text{Med} & \text{Avg} & \text{CoV} \\
  \midrule
  CP handlers   & 67843.636 & 77958.915 & 72613.399 & 72367.637 & \percentage{0.055} \\
  Null handlers & 71743.598 & 76642.726 & 72792.497 & 73450.008 & \percentage{0.025} \\
  No handlers   & 70857.059 & 75721.317 & 74950.524 & 74056.952 & \percentage{0.026} \\
  \bottomrule
  \multicolumn{6}{r}{(Unit: \si{\kilo\tuple\per\second}, except for CoV)} \\
  &&&&&
  \end{tabular}
  \label{tab:CPOverhead}
\end{table}
}

In this part, we determine the performance overhead of the \ac{CP}-instrumented
code and show that the yielded overhead is marginal compared to the
non-instrumented version.  We instrumented the \ac{PCh} in all actors of the
\ac{YhSB} and ran it on \ac{GCE} nodes. Second, we ran the same configuration
but with all handlers nullified.  Lastly, we ran a version where we removed all
instrumentation in the source code. To measure the performance of the three
configurations, we used \acp{FLh}, which log the first and the last tuples to
determine the overall processing time, and we used \acp{PCh} in sink actors to
count the number of processed tuples. As a result, in
Table~\ref{tab:CPOverhead}, the \ac{CP}-instrumented \ac{YhSB} runs at
\SI[round-mode=places,round-precision=1]{72.367637}{\kilo\tuple\per\second} on
average, while the non-instrumented version runs at
\SI[round-mode=places,round-precision=1]{74.056952}{\kilo\tuple\per\second}. We
measured that the \ac{CP}-instrumented code yields \percentage{0.022} overhead
on average and \percentage{0.031} overhead in the median.

\section{CP Deployment: Measuring of Latency and Throughput\label{s:exp}}

\subsection{Analysis of Tuple Processing Latency\label{ss:exp_latency}}

In this section, we demonstrate how we used \ac{CP} for the tuple latency
analysis of the \ac{YhSB} and explain the significance of \ac{CP}, not only for
the benchmark but for the inter-node time duration measurements across cloud
nodes.  We used the same cloud configuration for the \ac{YhSB} in
Section~\ref{val:yahoo}. In this experiment, we maximized the incoming rate of
streaming data by enabling all five source actor instances, which is different
from the setting we used for the validation, where we deployed a single
instance of source actor to enable the \ac{ReT} method for the validation.

The benchmarking application has \acp{bIDh} instrumented at the beginning and
end of the tuple computation in all actor instances. It allows measuring the
tuple computation time and the time durations of all tuple transfers between
any two actors. For a tuple that has traveled the entire benchmark topology
from DGSpout to CampaignProcessor, its end-to-end latency can be measured by
relating two \ac{TSC} values, one from DGSpout and one from CampaignProcessor;
likewise, the \ac{CP} method can measure actor-to-actor latency by making a
relation using \ac{TSC} values from two consecutive actors.

\emph{Tuple transfer latency} is a time duration of a tuple that has
transferred from one actor to another.  \emph{Tuple processing latency} -- the
total processing time of one tuple throughout the topology -- is a tool to
assess the performance of a big data streaming topology.  We measure tuple
transfer latency between two actors, and the two actors may or may not be
connected directly.  Tuple processing latency is a partial case of tuple
transfer latency where the two measured actors are starting and end actors.

\emph{Poison pills} enable measuring a time duration on unbounded data streams
across multiple nodes. It takes the following steps: a streaming application
launches and starts emitting tuples. \Acp{FLh} instrumented in the source actor
instances recognize the first arriving tuple and timestamp it as the start of
the measurement duration.  Second, the streaming application emits poison pills
to notice the end of a period, and the poison pills start filling up the
application topology.  Lastly, \acp{FLh} in the sink actor instances recognize
the first arriving poison pill and mark the ending time. Having a total order
of start and end times, we can calculate the length of the period using the
earliest starting time and the latest ending time.

By instrumenting \acp{FLh} and \acp{bIDh} in this benchmark, we measured that
the benchmarking topology in this experiment ran for
\SI[round-mode=places,round-precision=2]{90.65588296173877}{\second},
processing \num{63000000} tuples during the run.

The latency between any two actors can be measured using \acp{bIDh}
instrumented in every actor of the benchmarking topology. In each actor, two
\acp{bIDh} log a tuple~ID and a \ac{TSC} value at the start and the end of the
tuple computation for every incoming tuple. We can calculate the tuple transfer
latency using the \ac{TSC} values of two actors in which a tuple started and
ended. The selected actors may or may not be connected; the end-to-end latency
of a tuple in the target topology is the case when we choose \ac{TSC} values
from a source actor and a sink actor.

\begin{figure}[tbh]
  \centering
  \includegraphics[width=\linewidth,trim={0 0 0 0},clip=false]
  {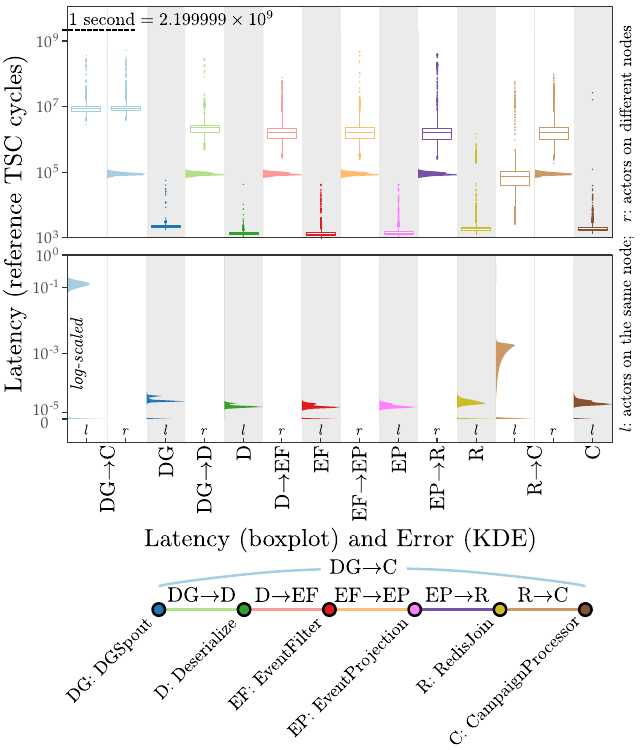} 
  \caption[Yahoo streaming benchmark latency analysis]
  {%
    The $x$-axis comprises individual actor-to-actor latency results from a run
    of \ac{YhSB}. For each latency result, a boxplot represents the
    distribution of latency, and a KDE represents the distribution of the
    \ac{CP} translation errors.
  }%
  \label{fig:latency_cloud}
\end{figure}

Fig.~\ref{fig:latency_cloud} shows the experimental results of the tuple
transfer latency between two actors.  The $y$-axis denotes the reference
\ac{TSC} cycles to depict latency and error, and the $x$-axis denotes either an
actor pair or an actor by itself; each actor pair shows inter-node latency
(tuple transferring time between two actors); each actor on the $x$-axis shows
tuple computation time.
For example, \emph{DG$\to$C} on the leftmost side of the $x$-axis is an actor
pair, and its boxplot shows tuple transfer time between the two actors DG and
C.  Next to \emph{DG$\to$C} is \emph{DG} -- its boxplot shows tuple computation
time on the actor DG.
different latency groups -- end-to-end (DG$\to$C), actor-to-actor (DG$\to$D,
D$\to$EF, EF$\to$EP, EP$\to$R, and R$\to$C).  Depending on the communication
distance between the measured actors, the latency result splits into
two sub-groups -- \emph{remote} and \emph{local}; the inter-node tuple transfer
latency as the remote group, and the node-local tuple transfer
latency as the local group.

We present the result using a boxplot. The width of a box represents the number
of latency-data points used to draw it. A black dot inside the box shows the
mean value of the latency group. A band within the box represents the median or
the \ordinalnum{2} quartile (Q2) (\textit{i.e.}, the \ordinalnum{50}
percentile).  The box's floor and ceiling denote the \ordinalnum{1} quartile
(Q1) and the \ordinalnum{3} quartile (Q3). The height of the box represents the
interquartile (IQR). The vertical lines connected to the above and the beneath
are whiskers, which represent the ranges $[\text{Q3},\text{Q3}+(1.5\times
\text{IQR})]$ and $[\text{Q1}-(1.5\times \text{IQR}),\text{Q1}]$, respectively.
Each whisker ends with a bar that denotes the maximum and minimum valuew and
the minimum value omitting outliers.  A group of vertically aligned dots above
and beneath a whisker is outliers. We present the error distribution of each
box as a kdernel density estimate (KDE) diagram. The remote groups yield a
maximum error of \CPINTERMAX{} and a minimum error of \CPINTERMIN{}. The local
groups yields a maximum error of \CPINTRAMAX{} and a minimum error of
\CPINTRAMIN{}. Intra-node durations incur errors because we translate them into
the reference \ac{TSC} cycles. The error results do not include time durations
on the reference node because they do not incur errors.

\begin{figure}[tbh]
  \centering
  \includegraphics[width=\linewidth,trim={0 0 0 0},clip=true]
  {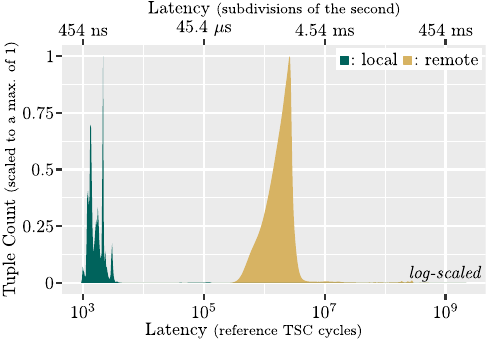}
  \caption[Actor-to-actor latency distribution]
  {%
    Actor-to-actor latency distribution using kernel density estimation.
    Latency of tuple computation time on the same node is aggregated closely to
    $10^3$~cycles.  A small portion of latency results close to $10^5$~cycles
    (green) represent tuple transfer on the same node whereas latency of
    inter-node-transfer (remote) is condensed closed to $10^6$~cycles.
  }%
  \label{fig:kde_cloud_e2e}
\end{figure}

\emph{End-to-end latency} is on the leftmost of Fig.~\ref{fig:latency_cloud}.
It measures the tuple transfer latency between DGSpout and CampaignProcessor.
Because it accumulates the tuple computation time of the intermediate actors
between the source actor and the end actor, it is also called \emph{end-to-end
processing latency} or tuple processing latency. The result splits into remote
and local.  Remote latency is the time duration between two actors that reside
on different nodes, and local latency, on the other hand, is the case where the
two actors reside on the same node.  Local latency, however, is not limited to
the case where tuples stay on the same node during the whole process; instead,
it includes a multi-node-traveling case where tuples begin and end on the same
node but move to other nodes during the processing. This case is effectively
depicted in Fig.~\ref{fig:kde_cloud_e2e} -- most of the tuple latency equal or
greater than \SI{1}{\milli\second} belongs to the remote latency.  However,
some cases in the local tuple latency are longer than \SI{1}{\milli\second}
because they belong to the multi-node-traveling case.  The diagram intuitively
differentiates end-to-end tuple processing latency from the local latency to
the remote latency -- the remote tuple transfer latency is mostly around
\SI{1}{\milli\second}, slower than the local tuple transfer latency.  The
interquartile of the remote end-to-end latency in Fig.~\ref{fig:latency_cloud}
also covers the \SI{1}{\milli\second} latency, which supports the result from
Fig.~\ref{fig:kde_cloud_e2e}. The result shows a glaring result that we cannot
measure inter-node time durations within the accuracy of \SI{1}{\milli\second}
with the \ac{NTP}-based measurement methodology, which has an error~bound of
\SI{1}{\milli\second} at the minimum.

\emph{Actor-to-actor latency} exhibits the tuple transfer latency of two
consecutive actors. The result exposes Storm's default orchestration strategy
prioritizes vertical scaling over horizontal scaling; \textit{i.e.}, the
actor-to-actor latency group does not have a remote group except for the
\emph{RedisJoin} to \emph{CampaignProcessor} group. It means that every tuple
that has started from a source actor on a specific node stays on the node until
it reaches \emph{RedisJoin}. For those local groups, tuple transfer latency
stays below \SI{100}{\micro\second}. We can also discover that the number of
tuples decreases as the tuples go through \emph{EventFilter}.

\section{Related Work\label{s:rel}}

Kieker~\cite{Kieker2008,Kieker2012} is a profiling framework for monitoring
concurrent or distributed software systems. It is suited for the situation when
the users want to monitor and figure out the performance leaks.  Except for
general monitoring for Java-based systems, Kieker supports various features,
including failure diagnosis and performance anomaly
detection~\cite{RABISER2017309}.

While both \ac{CP} and Kieker performance measurement tools, they have
different features. Kieker runs above JVM, while the \ac{CP} can run on native
\ac{C++} through \ac{JNI}. Thus, \ac{CP} can get performance measurement with
an overhead of nanoseconds through native calls, but Kieker takes microseconds
to get it with JVM overhead.  For example, it takes \SI{3.8}{\micro\second} to
call a single method on an enterprise server machine, including
instrumentation, collecting performance data, and writing them to a file
system~\cite{Kieker2012}.  On the other hand, a single call for \ac{CP}
buffered id handler takes less than \SI{100}{\nano\second}.  \Ac{CP} can get
the high-resolution time measurement using system calls, but Kieker cannot
since Java does not provide measurement API with nanosecond
granularity~\cite{Java15}.

Several researchers have discussed the performance of
Kieker~\cite{RABISER2017309,SPASSMeter,Kieker2008}.  In~\cite{Kieker2008}, the
authors of Kieker deployed Kieker in a real-world multi-user Java web
application, and they report that Kieker incurs \SI{9.3}{\percent} server-side
response time overhead.

Kieker can be used to monitor and analyze general big data applications. On the
other hand, \ac{CP} is an event-based performance profiler with various types
of handlers. It is specialized for streaming applications, synchronizing over
different nodes using \acp{TSC} to provide accurate inter-node time latency
between multiple nodes; moreover, \ac{CP} can measure the target system's
maximum sustainable throughput.

{\setlength\tabcolsep{3pt}
\renewcommand{\arraystretch}{1.3}
\begin{table}[h]
\caption[Single-thread performance of Kieker monitoring records]
{%
   The single-thread performance of Kieker monitoring records in latency and
   throughput
}%
\begin{center}
\begin{tabular}{ r r r r }
\toprule
\multicolumn{2}{r}{}&Latency&Throughput\\
\multicolumn{2}{c}{100M Iterations}
&\multicolumn{1}{c}{($\text{ns}$)}&\multicolumn{1}{c}{($\text{MT}/\text{s}$)}\\

\midrule
\multirow{2}{*}{\rotatebox[origin=c]{90}{Kieker}}
 & EmptyRecord     & \num{4653} & \fnum{0.214915} \\
 & TimestampRecord & \num{5004} & \fnum{0.199840} \\
\bottomrule
\end{tabular}
\end{center}
\label{tab:kieker}
\end{table}
}

As shown in Table~\ref{tab:kieker}, we check the single thread performance of
Kieker monitoring records on \ac{GCE}.  While the handlers of \ac{CP} can
process at least several million tuples per second as shown in
Table~\ref{tab:cp}, the maximum throughput that Kieker can get is around
\SI{0.2}{\mega\tuple\per\second}.

Both the authors of~\cite{YahooBenchmark2016} and~\cite{Karimov18} use \ac{NTP}
to synchronize the time among multiple nodes; however, Dongen~\textit{et al.}~
\cite{Eval2020} note that \ac{NTP} is only accurate to a minimum error of
\SI{35}{\milli\second} for synchronization \cite{NTP35ms}.  They propose an
approach similar to \cite{ReTMethod}, where they use a single Kafka broker to
improve synchronization accuracy up to the millisecond level; however, in their
sustainable throughput measurement, they used a multi broker Kafka cluster
which relies on \ac{NTP} because a single Kafka broker cannot emit high
throughput to measure sustainable throughput of \acp{SPE}.  To increase the
accuracy of latency measurements, we propose \ac{CP} method which has an error
bound equal to the \ac{MinRTT} between two nodes.

Ridoux and Veitch present the software-based
\emph{TSCclock}~\cite{TenMicroseconds,RobustIMC2004,RobustJournal}, which
performs between the sub-microsecond accurate \ac{PTP} and the one millisecond
accurate \ac{NTP} for measuring time durations on a single computer; however,
the \emph{TSCclock} method has not been designed for the measurement of
inter-node durations for performance profiling and has the following
disadvantages compared to the proposed approach.
First, their approach introduces clocks, establishing a notion of global time.
The conversion of \ac{TSC} cycles to seconds requires the \ac{TSC} frequency,
or in other words, $\hat{p}$, an estimate of the average CPU oscillator period
$p$.  A typical quartz yields an error of 20~microseconds~per~second due to
thermal changes~\cite{Najafi2021}, and incorporating an oscillator period
increases the error above the required accuracy.
Second, to measure the time duration between a start event and an end event,
the 10-microsecond accurate difference clock $\mathit{C_d(t)}$ requires the two
events to be from the same \ac{TSC} so that the initial offset \emph{K} in both
timestamps cancel each other, \textit{i.e.}, the difference clock is incapable
of measuring the inter-node time duration between two \acp{TSC} because the
two timestamps have different \emph{K}s.
Third, the work does not provide a strict error bound on the obtained accuracy
of the $\mathit{C_d(t)}$ and $\mathit{C_a(t)}$ clocks.  They provide
statistical accuracy with a median and an IQR of their error. Fig.~3
of~\cite{TenMicroseconds} shows an error value higher than
\SI{50}{\micro\second} at specific instants, larger than the median
of~\SI{16.7}{\micro\second}, and an IQR of~\SI{7.8}{\micro\second}. It means
that the error of inter-node time duration measurements between two nodes can
be as high as \SI{100}{\micro\second}; likewise, their absolute clock
$\mathit{C_a(t)}$ has $\hat{\theta}$, the current estimate of the uncorrected
clock. Its median error is \SI{30}{\micro\second}, which is not
bounded~\cite{RobustIMC2004,RobustJournal}. When measuring the inter-node
duration between two nodes, the errors of the two uncorrected clocks add up.
Lastly, the proposed approach requires kernel modifications to timestamp
incoming and outgoing UDP packets. Modifying the host kernel from the guest~OS
is infeasible in hypervised environments such as public clouds.

The timestamp-counter-based precise Relative Clock Synchronization Protocol
(TSC-RCSP)~\cite{Tian2014}  achieves a 10-microsecond accuracy for measuring
the relative offset between clocks. It uses \ac{TSC} values and the \ac{TSC}
frequency to measure the time offset between two clocks. They calculate the
frequency every \SI{100}{\second} to mitigate the problem with the oscillator
frequency error. TSC-RCSP has the following shortcomings.
First, the protocol requires a modified network interface card (NIC) to acquire
timestamps. They developed a kernel module to calculate the \ac{TSC} frequency
periodically; however, such a modification is infeasible for the guest OS in a
hypervised environment.
Second, the error of TSC-RCSP is not bounded. TSC-RCSP guarantees a
synchronization accuracy of \SI{10}{\micro\second} with \SI{97}{\percent}
confidence.
Third, TSC-RCSP does not target general-purpose communication networks,
\textit{e.g.}, LAN or the Internet. Instead, the authors designed the protocol
for distributed real-time control systems, where real-time support is assumed,
and fixed-size packets travel via a control network.

Chen and Yang propose a clock synchronization scheme using a Time
Synchronization Function (TSF) counter available in Wi-Fi network
cards~\cite{ptpwifi2021}. The authors emulate a hardware \ac{PTP} clock using
the TSF counter and show that they achieve \SI{1}{\micro\second}-level accuracy
between Wi-Fi devices.

The authors of~\cite{OneWay} describe the algorithm that compares the accuracy
of the \emph{One-Way Delay (OWD)} estimation mechanisms.  OWD is a formal
estimation of half the \ac{RTT}, \textit{i.e.}, $\left(\text{RTT} \cdot
0.5\right)$.  This informal estimation \emph{average (av)} can not be the true
OWD because network time always differs on the way to a target node and the way
back to the originated node.  In this work, the authors propose a way to
formally compare their previous contribution \emph{minimum pairs (mp)} to the
informal estimation algorithm.

\emph{Time-of-Flight Aware Time Synchronization Protocol
(TATS)}~\cite{Thiele:2016} is a multi-hop time synchronization protocol for
energy-consumption-aware wireless devices which achieves sub-microsecond
synchronization error. It tackles the propagation delay problem between
antennas by applying compensation to each link with an estimated delay shipped
on a message.

Researchers have observed that \acp{SPE} start to exhibit backpressure when the
tuple arrival rate exceeds the processing capacity of the SPE.  A widely used
data generator, Apache Kafka, suffers from low throughput, capped at
\SI{420}{\kilo\tuple/\second}~\cite{Kafka:eval}. The backpressure issue
requires well-defined criteria for how much throughput the streaming frameworks
can process without backpressure.

Karimov~\textit{et al.}~\cite{Karimov18} newly defined and measured,
sustainable throughput that previous work, such as the
\ac{YhSB}~\cite{YahooBenchmark2016}, omitted.  The authors define
\emph{sustainable throughput} as the highest load of event traffic that a
system can handle without exhibiting prolonged backpressure.  Dongen~\textit{et
al.}~\cite{Eval2020} further extend the notion of sustainable throughput by
defining maximum threshold values for latency and CPU utilization.

Both works \cite{YahooBenchmark2016} and \cite{Eval2020} used Apache Kafka as a
data generator.  A single data generator of \cite{YahooBenchmark2016} falls
behind at around 17,000 events per second.  \cite{Karimov18} mentions that
Apache Kafka and Redis are the bottlenecks of the Yahoo! Streaming
Benchmark~\cite{YahooBottleneck}.  They choose to generate data on-the-fly and
eliminate fluctuations due to network issues and GC interruptions.

A large body of work~\cite{YahooBenchmark2016, Eval2020, StreamBench,
PerfComp2016, CompFrame2017, Senska, nasiri_evaluation_2019, ExperiEval,
DSPBench, NAMB, Clickstream2021} used Apache Kafka as a data generation
facility for benchmarking streaming frameworks, as it is a widely used message
broker. It was pointed out in~\cite{StreamBench} that Kafka could be a
bottleneck due to their data generation speed and messaging processing
capability. Researchers proposed workarounds to mitigate those limitations of
Apache Kafka. To ensure that Apache Kafka does not degrade the performance,
Lu~\textit{et al.}~\cite{StreamBench} preloaded and replayed a set of records
during the experiment to avoid the data creation overhead.
Nasiri~\textit{et al.}~\cite{nasiri_evaluation_2019} ran Kafka in a separate
machine.  Karimov~\textit{et al.}~\cite{Karimov18} abandoned Kafka and created
new data generators to overcome the bottleneck and generate the data on
the fly.  Shukla~\textit{et al.}~\cite{RIoTBench} implemented a scalable
data-parallel event generator that acts as a source task for IoT sensors at a
throughput of \SI{10}{\kilo\tuple\per\second}.  Neither~\cite{Karimov18}
nor~\cite{RIoTBench} have our proposed object factory to avoid data
deserialization overhead, and they do not qualify our strict definition of
sustainable throughput.

Pal~\textit{et al.}~\cite{Clickstream2021} proposed a clickstream analysis for
an e-commerce system.  Their contribution includes a stream processing analysis
that implements operations at different complexity levels: simple and
low-latency stream processing operations and time-consuming data mining
operations.  With their experimental result, they conclude that real-time
workloads prefer stream processing over batch processing.
Their streaming platform is a part of their big data Lambda architecture that
consists of Apache Hadoop and Apache Storm as data processing frameworks and
Apache Kafka as a data generator. They ran the experiment on Microsoft Azure
HDInsight cloud, deploying four Storm worker nodes, two Storm Nimbus nodes, and
three Zookeeper nodes.

\section{Conclusions\label{sec:concl}}

Existing big data stream processing engines cannot ingest millions of
events-per-second incoming to the system. Our performance engineering efforts
to improve the current streaming engines' low throughput and high latency have
refurbished the entire framework in different dimensions. First, we have
proposed the \ac{CP} method, a profiling technique to measure time durations
between two nodes by relating \acfp{TSC} of cloud nodes. \Ac{CP} has improved
the measurement accuracy up to \CPINTERMAX{}, which is three orders of
magnitude higher than the accuracy of \ac{NTP}, and an order of magnitude
higher than prior work. Second, we have devised and demonstrated a
throughput-controlled data generator to determine the sustainable throughput of
a streaming engine. Using our data generator, we have shown that the
sustainable throughput of the Apache Storm framework running on the Google
Compute~Engine has increased from \SI{700}{\kilo{}} to \SI{4.68}{\mega{}}
tuples per second. Third, we have presented a concurrent object factory that
facilitates our data generator by eliminating the deserialization overhead
incurred by a stream processing engine.

\ifCLASSOPTIONcompsoc
  \section*{Acknowledgments}
\else
  \section*{Acknowledgment}
\fi
This research has been supported by the Next-Generation Information Computing
Development Program through the National Research Foundation of Korea (NRF),
funded by the Ministry of Science \& ICT under grant
No.~NRF\-2015\-M3C4A\-7065522, by the Global Ph.D.\ Fellowship Program through
the NRF funded by the Ministry of Education under Grant
No.~NRF\-2015\-H1A2A\-1033965, and by the NRF funded by the Korean government
(MSIT) under Grant No.~2019\-R1F1A\-1062576.

\ifCLASSOPTIONcaptionsoff
  \newpage
\fi

\bibliographystyle{IEEEtranUrldate}
\bibliography{references}

\end{document}